\documentclass{jfm}
\usepackage{xcolor}
\usepackage[draft]{todonotes}
\usepackage[round,super]{natbib}
\usepackage[colorlinks=true, linkcolor=blue, citecolor=blue, bookmarks=true]{hyperref}
\usepackage{changebar}

\usepackage{epstopdf}
\usepackage{amsmath}
\usepackage{amssymb}
\usepackage{graphicx}
\usepackage{color}
\usepackage{textcomp}
\usepackage{epsfig}
\usepackage[applemac]{inputenc}
\usepackage{natbib}
\usepackage{array}
\usepackage{bm}
\usepackage{setspace}
\usepackage{latexsym}
\usepackage{tikz}
\usetikzlibrary{shapes.geometric}
\usetikzlibrary{arrows.meta,arrows}

\shorttitle{Vorticity Production at  Fluid Interfaces in Two-dimensional Flows  }
\shortauthor{D. Fuster and M.Rossi}
\title{Vorticity Production at  Fluid Interfaces in Two-dimensional Flows }

\author{
Maurice Rossi $^{1,2}$
   \and
Daniel Fuster $^{1,2}$
}
\affiliation{$^1$ Sorbonne Universit\'e, UPMC Univ Paris 06, UMR 7190,
     Institut Jean Le Rond d'Alembert, F-75005, Paris, France \\[\affilskip]
    $^2$ CNRS, UMR 7190, Institut Jean Le Rond d'Alembert,
     F-75005, Paris, France
}

\begin{document}
\maketitle

\begin{abstract}
This work revisits the  production of vorticity at  an interface separating two immiscible incompressible fluids. A new decomposition of   the vorticity flux is proposed  in a  two-dimensional context  which allows to  compute    explicitly     such a  quantity in terms of      surface tension $\sigma$,  viscosity $\mu$ and gravity $g$.  This approach is then applied   in the context of  gravito-capillary waves. It leads to analytical results already known but from  a new perspective,  provides some quantitative predictions at short time that can be a good test for numerical codes. Finally it is a mean to obtain a qualitative understanding  of    direct numerical  simulation results. 
 \end{abstract}

\section{ Introduction  }
 \label{introductionsection}

\noindent   When  fluid  properties  such as density   are  varying   in a spatial  domain, vorticity is  generated in that domain   by a  baroclinic effect  \cite{magnaudet2020particles}.  If this region is  an extremely thin layer, it is considered from the view point  of continuum mechanics  to be a  sharp discontinuity between two fluids.  As a consequence,  fluid/fluid interface plays  the role of  a source of vorticity as for a fluid/solid interfaces. Contrary to the fluid/solid interface,  the produced vorticity  eventually influences the dynamical response of the interface itself. This manuscript   discusses  and quantifies the generation of  vorticity produced at the interface between two incompressible and immiscible fluids within the context of  two-dimensional flows. 

\vspace{0.2cm}
   
\noindent Initial theoretical works on vorticity field at  interfaces  were devoted to the particular case of a free surface flow,  imposing zero shear stress at the interface and a constant pressure  in the outer fluid due to the neglect of  the outer fluid dynamical viscosity and density.  Various papers \cite{longuet1953mass,longuet1992capillary} obtained  in a steady two dimensional flow, 
(see also \cite{batchelor2000introduction}),  the  relation $\omega = - 2\kappa ~(\vec{u} \cdot \vec{t})$
giving  vorticity at   free boundary $\omega$ as a function of  interface curvature $\kappa$  and  tangential velocity  $\vec{u} \cdot \vec{t}$. 
This result and  its generalization to three-dimensions \citep{longuet1998vorticity,peck1998kinetics},
stress the intrinsic relation between  interface topology and  vorticity intensity. 
 Furthermore the presence of thin vorticity layers  at the interface have motivated the development of theoretical and numerical methods based on the boundary-layer approach where the potential flow far from the surface is constrained by the conditions imposed by an infinitely thin boundary layer at the interface.  These models have  successfully  reproduced the response of weakly damped Stokes waves \citep{longuet1992capillary} and  obtained constraints in the relation between flow properties and vorticity field at the interface in two-dimensional steady flows  imposed by the stress free condition \citep{sarpkaya1996vorticity}. 

\vspace{0.2cm}

\noindent The  relation  $\omega = - 2\kappa ~(\vec{u} \cdot \vec{t})$   is based on a kinematical relationship plus the definition of zero shear but does not depend on  the condition imposed on pressure at the interface.  This is related to the fact that, vorticity produced at the surface   diffuses into the bulk  \citep{longuet1953mass,longuet1960mass} although
this process is not given by the previous relation. More precisely,    the value of vorticity  imposed at the free surface  does not  provide the rate at which vorticity diffuses into the  bulk that ultimately is 
the vorticity production rate. For example while the vorticity along a steady flat surface  is null, the vorticity production
and the diffusion of vorticity towards the interface is found to be proportional to
the pressure gradient along the surface and therefore not necessarily null in this particular case \citep{lugt1987local}. 
Thus,  following the work of \citet{lighthill1963boundary} where
a vorticity sheet along the interface is introduced to satisfy
the non-slip boundary condition in a fluid/solid interface,
various authors have focused their efforts to derive  generalized expressions for the vorticity generation at a free surface in non-steady flows equations   \citep{rood1994interpreting,lundgren1999generation}.

  \vspace{0.2cm}
 
\noindent The initial works on the production of vorticity by interfaces in free surface flows have been
extended to interfaces between two immiscible fluids with arbitrary values of density and viscosity
allowing to study the influence of vorticity production beyond the limiting case of free surface flows.
\citet{lugt1987local} already discusses some constrains imposed in the angle between streamlines at the surface
between two viscous fluids.  The dynamic viscosity ratio   was identified to be the 
parameter controling the streamline patterns for steady two-dimensional incompressible flows.
Later, \cite{wu1995theory,wu1998boundary} expressed  the production of vorticity by sharp interfaces between two viscous fluids  and \cite{dopazo2000vorticity} establishes the relations that must be used
in vorticity-based formulations for the interface between two viscous fluids 
providing also an evolution equation for the vortex sheet-strength.
More recently \cite{brons2014vorticity,brons2020vorticity} and \cite{terrington2020generation}
have extended the work of \cite{lundgren1999generation} to obtain general expressions for the production
of vorticity by fluid/fluid interfaces in two--dimensional flows for both non-slip and stress-free conditions 
discussing the consequences of vorticity generation in various problems involving the presence of interfaces between
two viscous fluids.

  \vspace{0.2cm}
 
\noindent  In the present paper, we revise the previous  expressions of the vorticity production
in two--dimensions. First  we  provide a symmetric expression for  the sources of vorticity. Second and more importantly, we derive
the explicit dependency of  the vorticity production with respect to surface tension, viscosity and gravity. 
Thereafter we apply this approach   to the problem of   
gravity--capillary waves widely studied in the literature
for both linear \citep{lamb1945hydrodynamics,prosperetti1981motion}
and non-linear regimes \citep{lundgren1989free,fedorov1998nonlinear}.
It is shown  that this  formulation is useful to understand
the transition between the linear and non-linear regime and explain the symmetry  breaking of 
  oscillations of gravity waves.

 \vspace{0.2cm}
  
 \noindent The manuscript is structured as follows. Section~\ref{VorticiyProductionInterfaceGENERAL2D}  recalls some previous theoretical results about  vorticity production at an interface. Section \ref{VorticiyProductionInterfaceGENERAL2DSYMMETRIC}  introduces a  new    decomposition of the production term  in terms of pressure jump and mean pressure at  interface.  The procedure to compute the mean pressure   is then  presented. The result of this approach is an  explicit expression of vorticity production which discriminates  the  various mechanisms (section~\ref{vorticityproductionexpression}).  In the end of this section   some asymptotic cases are given.   Such a theoretical    result  is then applied on   the flow example  of a gravito-capillary wave. For linear   viscous waves, one recovers previously known analytical  results in an alternative way (section~ \ref{GravitocapillaryAnaly}).   For the non-linear regime,  cases of non-linear  evolution are  studied from this  new viewpoint,  providing some qualitative  interpretations (section \ref{bumpflowgrav}).

\section{Vorticity production   at  interfaces.}
\label{VorticiyProductionInterfaceGENERAL2D}

\noindent   This section recalls the main previous theoretical results and defines our notations. Let us  consider a  two-dimensional   flow  with  two incompressible and
immiscible fluids  separated by an interface   $(I)$  possessing a      
surface tension  $\sigma$.  Each fluid   \footnote{Whenever
necessary,  the notation  $Q^{(r)}$  is  explicitly used to  stand for  the
quantity $Q$ in phase $r=1$ or $r=2$.   Notations $[[Q]]$ and $Q_m$  respectively  stand for the difference
$(Q^{(1)}-Q^{(2)})$ and the mean value $(Q^{(1)}+Q^{(2)})/2$ at a point of the interface.}  possesses  a   constant   density $\rho^{(r)}$  or equivalently  a specific volume $\upsilon^{(r)} \equiv  {1}/{\rho^{(r)}} $, a  constant   dynamic viscosity $\mu^{(r)}$  or equivalently a  kinematic viscosity $\nu^{(r)}\equiv  \mu^{(r)} \upsilon^{(r)} $.  In  the    plane $(x,y)$ oriented by the unit normal $\vec {e}_z$,  interface   position  and    velocity  components $u_i(x,y,t), ~~i=1,2$     provide a complete description of    the flow (we use in  some obvious occasions  $x$ for $x_1$ and $y$ for $x_2$).  This flow  is  governed by  the  Navier--Stokes  equation
\begin{equation}
\rho^{(r)} \frac{ D {u}_i^{(r)}}{D t}   =\partial_j   \tau_{ij}^{(r)}   +  \rho^{(r)} F_i(\vec{x},t),~~~~~\partial_i u_i=0 
\label{Meqn1:999}
\end{equation}
where $D/Dt \equiv \partial / \partial t + {u}_i^{(r)} \partial_i   $ denotes  the Lagrangian time derivative and  $\tau_{ij}^{(r)}  $ the stress tensor 
\begin{equation}
   \tau^{(r)}_{ij}= -p^{(r)}\delta_{ij} +2 \mu^{(r)} e^{(r)}_{ij},~~~~~~e^{(r)}_{ij}=\frac{1}{2}[\partial_i u^{(r)}_j+\partial_j u^{(r)}_i].
\label{Meqn1STRESSTENSOR}
\end{equation}
 The force   \textit{per  unit mass}    $\vec{F}(\vec{x},t)$     is assumed to be  continuous across the interface and to derive from a potential i.e.  $\vec{F}=-\vec{\nabla}[g \varphi]$ where $g$ is an amplitude parameter.  In the example studied below, $\vec{F}(\vec{x},t)$  is gravity :  in that instance,  coordinate  $y$ stands for  the     upward vertical cartesian coordinate and  $\varphi=  y$  and  $g= 9.81$  the acceleration of gravity  in S.I. units.

  \vspace{0.5cm}

\noindent   Vorticity field is directed along  $\vec {e}_z$ and 
 its    unique component    $\omega(x,y,t)$  is  governed by 
 \begin{equation}
\frac{D\omega^{(r)}}{D t}  =    - \nabla  \cdot  \vec{J}^{(r)} , ~~~ \vec{J}^{(r)}  \equiv - \nu^{(r)} \, \vec{\nabla}\omega^{(r)}
\label{MeqVorticity}
  \end{equation} 
At any point of   interface $(I)$, one    attaches  a   Frenet--Serret   frame.
One may select  the unit normal vector $ \vec{n}^{i \to o }$  directed from phase $i$ to phase $o$ (two cases are possible $(i,o)=(1,2)$ or $(i,o)=(2,1)$) and  the tangent  vector $\vec{t}^{i \to o} \equiv \vec{e}_z \times  \vec{n}^{i \to o}$. The portion of interface  $(I)$    is oriented  so that the curvilinear abscissa $s$ is  directed along  $\vec{t}^{i \to o}$.  As a consequence,  
\begin{equation}
\vec{t}^{i \to o} \equiv \frac{\text{d}\vec{x}}{\text{d}s},~~~~ \frac{\text{d}\vec{t}^{i \to o}}{\text{d}s} \equiv   \kappa \vec{n}^{i \to o},~~~~~~~ \frac{\text{d}\vec{n}^{i \to o}}{\text{d}s} \equiv  -  \kappa \vec{t}^{i \to o} 
\label{MeqnserretFrenetbasis12D}
\end{equation}
in which   the radius of curvature        $R \equiv 1/ \kappa $   is  positive when the curvature  center is in region $o$. Across the interface,  the velocity field is continuous and  the Young--Laplace law is enforced \citep{batchelor2000introduction}. These constrains impose   several conditions  on velocity gradient tensor $\partial_i  u_j$,    vorticity components and pressure: 
\begin{equation}
{t}^{i \to o}_i \partial_i  [[  u_j ]]    =0,~~~~~~n^{i \to o}_i \left[\left[\partial_i  u_j \right] \right] n^{i \to o}_j =0,~~~~~[[\omega ]] = [[\frac{ 1}{  \mu}]]  {n}^{i \to o}_i  \tau_{ij}  {t}^{i \to o}_j. 
\label{MeqomegaRjump}
\end{equation}  
\begin{equation}
 [[p]] = -  \frac{\sigma}{R}  {n}^{i \to o}_j  {n}^{1 \to 2}_j + 2\,[[\mu]] \, u_j \frac{ {n}^{i \to o}_j }{R}   -2\,[[\mu]] \,     {t}^{i \to o}_i \partial_i (u_j {t}^{i \to o}_j).  
\label{MeqPRESSUREDROP}
\end{equation} 
 
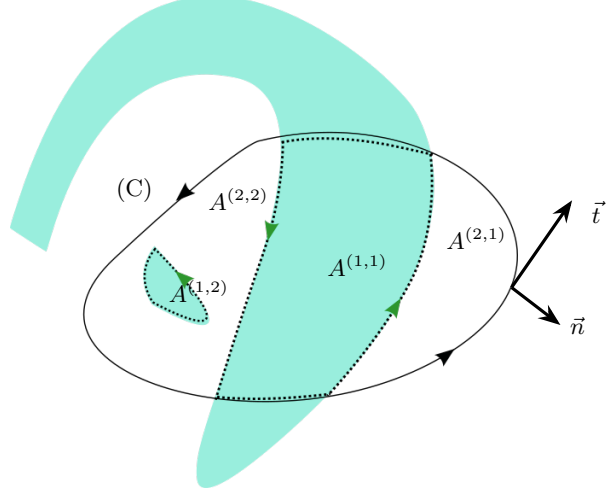
\begin{figure}
\begin{picture}(200,200)
  \put(110,0){\includegraphics[width=0.50\textwidth]{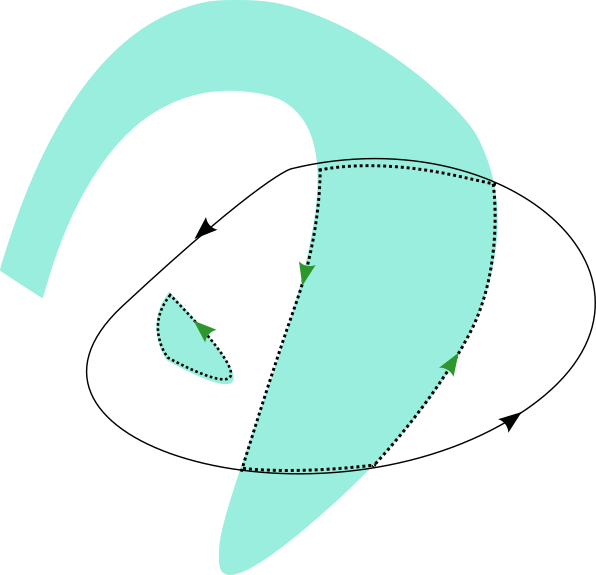}}
  \put(150,110){(C)}
  \put(330,100){$\vec{t}$}
  \put(230,80){$A^{(1,1)}$}
  \put(275,90){$A^{(2,1)}$}
  \put(170,70){$A^{(1,2)}$}
  \put(185,105){$A^{(2,2)}$}
\begin{tikzpicture}[
            > = Straight Barb,
phasor/.style = {very thick,-{Stealth}},
angles/.style = {draw, <->, angle eccentricity=1,
                 right, angle radius=7mm}
                        ]
\put(299,55){\draw[phasor] (21,13) -- ( 30:25) coordinate (v)  node[right] {$\vec{n}$};}
\put(299,55){\draw[phasor] (21,13) -- ( 33:26)  node[left] {};}
\end{tikzpicture}
\end{picture}
\caption{  A typical geometry.}  
\label{figScheme}
\end{figure}


   \vspace{0.5cm}

\noindent   Consider a  surface $A$   delimited by  a  closed Lagrangian curve $(C)$ (figure \eqref{figScheme}).  Surface $A$ is divided   in $A^{(1)}$ and $A^{(2)}$  belonging respectively  to  phase $k=1$ and $2$. Each $A^{(r)}$ is a set   of connected  surfaces    $A^{(r,p)}$. Four types of   $A^{(r,p)}$ are possible. First  a simply connected  surface  delimited  by  a closed curve entirely  part of   interface  $(I)$, second   a simply connected  surface delimited  by   one or two    open curves of  interface    $(I)$    ending  on $(C)$ plus  one or two  curves belonging to  $(C)$ which can be reduced to a unique point.    Third and fourth    cases are similar to the first or second  cases  except that  they are not simply connected but  contain  one or several  holes of the opposite phase as well. As far as orientation is concerned, we use always the outwards unit  normal vector: that is    $ \vec{n}^{1\to 2}$ and the  tangent unit  vector  $\vec{t}^{1\to 2}= \vec{e}_z  \times  \vec{n}^{1\to 2}$ for set $A^{(1,p)}$    (resp.  $ \vec{n}^{2\to 1}$ and $\vec{t}^{2\to 1}$  for set $A^{(2,p)}$). For closed curves, this completely defines the set which is compatible with the chosen orientation of  $(C)$.    
  Let us now  integrate equation \eqref{MeqVorticity} on  the two monophasic  regions $A^{(1)}$ and  $A^{(2)}$ , it is easily seen that  
 \begin{equation}
\frac{\text{d}}{\text{d}t}\left(\int_{A^{(1)}}\omega\,\text{d}x  \text{d}y\right)
  =    -\int_{(C_1) } J^{(1)}_{j}\,n_{j}\,\text{d}s_c + \int_{(I) }  \Sigma^{(1)}\,\text{d}s,
  \label{MeqnIntegomegaJijV1}
\end{equation}
 \begin{equation}
\frac{\text{d}}{\text{d}t}\left(\int_{A^{(2)}}\omega\,\text{d}x  \text{d}y\right)
  =    -\int_{(C_2)} J^{(2)}_{j}\,n_{j}\,\text{d}s_c + \int_{(I)}  \Sigma^{(2)}\,\text{d}s,
  \label{MeqnIntegomegaJijV2}
\end{equation}
where quantity $\Sigma^{(1)}$ (resp. $\Sigma^{(2)}$ ) plays the role of a source term for vorticity production in  phase 1 (resp.2) on interface $(I)$ 
 \begin{equation}
 \Sigma^{(1)} \equiv - n^{1\to 2}_{j} \,J^{(1)}_{j},~~~ \Sigma^{(2)} \equiv   n^{1\to 2}_{j}\, J^{(2)}_{j}
   \label{MeqnIntegomegaJijVsigma}
\end{equation}
and $(C_1)$ or  $(C_2) $ is the part of contour $(C)$ in phase 1 or 2. Vector $\vec{n}$ denotes  the outgoing unit  normal vector on  curve $(C)$ and tangent vector $\vec{t} = \vec{e}_z \times \vec{n}$ prescribes the orientation of   $(C)$ and hence the curvilinear coordinate  $s_c$   that    increases in the direction of $\vec{t}$. Variable  $s$  is  the curvilinear coordinate   defined on $(I)$    that    increases in the direction of $\vec{t}^{1\to 2}$.  Vorticity sources can be written  based on the obvious equality between the viscous term in Navier-Stokes equation and the vorticity flux~:
 \begin{equation}
 \nu^{(r)}\Delta \vec{u}^{(r)}= - \vec{e}_z  \times  \vec{J}^{(r)} 
\label{NSequationvectorielle}
  \end{equation}
or rather its projection    on the tangent  vector $\vec{t}^{1\to 2}$  on $(I)$
 \begin{equation}
 {t}^{1\to 2} \cdot   [   \nu^{(r)}\Delta \vec{u}^{(r)}]=  - (\vec{t}^{1\to 2} \times \vec{e}_z)  \cdot \vec{J}^{(r)}= -\vec{n}^{1\to 2}\cdot \vec{J}^{(r)},  
  \label{NStangentBIS}
  \end{equation}
  The source $\Sigma^{(1)}$ in phase 1  and   source $\Sigma^{(2)}$ in phase 2 are hence
\begin{equation}
 \Sigma^{(1)}  \,=\nu^{(1)}  {t}^{1\to 2}_i  \partial_j\partial_j {u}^{(1)}_i,~~~ \Sigma^{(2)}  \,= -\nu^{(2)}  {t}^{1\to 2}_i  \partial_j\partial_j {u}^{(2)}_i 
  \label{MeqnIntegomegaJijVsigma}
\end{equation}  
 Using   the  Navier--Stokes  equation, alternative expressions
   \begin{equation}
\Sigma^{(1)}=    \frac{\partial  }{\partial s}  (\frac{p^{(1)}}{\rho^{(1)} } )   +  \vec{t}^{1\to 2} \cdot [ \frac{D\vec{u}}{Dt}  - \vec{F}],~~~~
\Sigma^{(2)}=  - \bigg(\frac{\partial  }{\partial s} (\frac{p^{(2)}}{\rho^{(2)} })   +     \vec{t}^{1 \to 2} \cdot  [ \frac{D\vec{u}}{Dt}  - \vec{F}]    \bigg)
 \label{NStangentSources}
 \end{equation}
where   $\frac{\partial Q}{\partial s}$ stands for  the derivative $\vec{t}^{1\to 2} \cdot \vec{\nabla} Q$  of any quantity $Q$ defined on  interface  $(I)$.

\noindent Summing the two integrals \eqref{MeqnIntegomegaJijV1}-\eqref{MeqnIntegomegaJijV2} yields the dynamics of the  flux of vorticity over a Lagrangian   surface $A$ delimited by  a  closed Lagrangian curve $(C)$   
 \begin{equation}
\frac{\text{d}}{\text{d}t}\left(\int_{A}\omega\,\,\text{d}x  \text{d}y\right)
  =     -\int_{C} J_{j}\,n_{j}\,\text{d}s_c
   +\int_{I} \Sigma \,\text{d}s,~~~~~\Sigma \equiv \Sigma^{(1)}+ \Sigma^{(2)}.
   \label{MeqnIntegomegaJijVtotbis}
 \end{equation}
 The first r.h.s.  integral is a classical viscous diffusion term  through a boundary  $(C)$ and the second integral is a supplementary term corresponding  to the vorticity sources located at the interface.  Because of no-slip  condition  must be satisfied at any time on interface  $(I)$,  accelerations are continuous across $(I)$ 
  \begin{equation}
\frac{Du^{(1)}_i}{Dt}=\frac{Du^{(2)}_i}{Dt}
\label{MeqnAcceleration}
 \end{equation}  
 The above   equation together with \eqref{NStangentSources} yields the total vorticity flux 
  \begin{equation}
\Sigma \equiv \Sigma^{(1)}+ \Sigma^{(2)}= \frac{\partial  \Psi_{\Sigma} }{\partial s}~~~~~~\Psi_{\Sigma} \equiv [[\upsilon {p}]]
\label{MeqnIntegomegaJijVtot}
 \end{equation}
Vorticity   is   generated on  interface $(I)$   to comply with  the constraints on  the velocity field,  otherwise stated  it is  related  to the  boundary layer  present close to  an interface.   The  sources of vorticity are always generated by the scalar products of $\vec{t}^{1\to 2}$ with a gradient. When  interface $(I)$ is a closed contour (e.g. phase 1  included inside phase 2), the total production of vorticity $ \int_{(I)}  \Sigma\,\text{d}s$ is hence null: negative and positive production are opposite.  The previous  results   were already presented  in 
\cite{brons2014vorticity} as well as in \cite{wu1995theory}.

  \vspace{0.5cm}

\section{A symmetric decomposition of Vorticity production.}
\label{VorticiyProductionInterfaceGENERAL2DSYMMETRIC}

  \noindent In  \cite{brons2014vorticity},  the trivial identity 
  $[[AB]]=A^{(1)}[[B]]+ [[A]] B^{(2)}$ was used with $A=\upsilon  $  and $B =p$, to split  the source in two terms 
  $[[\upsilon {p}]] =   \upsilon^{(1)}  [[p]] + [[\upsilon ]] p^{(2)}$.  A first  term is      the jump in pressure which is known from Laplace law as a function of parameters such as $\sigma$, $\mu$ and  $g$.  This decomposition however   yields an asymmetric  treatment  of  the two phases. In addition the   second term  i.e. the pressure in phase 2 contains  implicit    dependencies on  parameters. 
 In the present paper,  
 a   symmetric identity  is used   instead   and   dependencies on the
 parameters are made  explicit.  First  the  force density
 $\vec{F}=-\vec{\nabla}[g\varphi]$   is  continuous across the interface and
 can be thus  included in a new  pressure term 
 \begin{equation}
   P^{'(r)} \equiv    p^{(r)} +  \rho^{(r)}g \varphi.
 \end{equation}
 For   gravity, the interface is located at  $y=\eta(s,t)$;   $y$ being the upward vertical then $\varphi(s,t) =
 \eta(s,t)$.   This is a way to split the action of hydrostatic pressure from
 the remaining contributions of   pressure. Second  using the symmetric
 identity 
 \begin{equation}
[[ AB]]=A_m[[B]] + [[A]]B_m.
 \label{diffpsurrho}
 \end{equation}    
  with   $A=\upsilon$ and $B= P'$, the source term  \eqref{MeqnIntegomegaJijVtot} becomes  
\begin{equation}
  \Sigma   =  \upsilon_m  \frac{\partial [[P']]}{\partial s} +[[\upsilon]]  \frac{\partial P'_m}{\partial s}    
  \label{SigmaALternativeNew} 
 \end{equation}   

This formulation    clearly separates  different effects.  On the one hand, the  first   term  contains  the pressure jump $[[P']]$    given by     Young -Laplace law   
 \begin{equation}
  [[P']] = - \sigma \kappa  +     g [[\rho]] \eta   + \,2\,[[\mu]] \bigg( \kappa \vec{u} \cdot \vec{n}^{1\to 2}
  - \, \frac{\partial }{\partial s} (\vec{u} \cdot \vec{t}^{1\to 2} )\bigg),  
 \label{MeqnrYoungLaplace2DNormalBIS}
\end{equation}  
where  $ y=\eta(s,t)$ is the interface position. This introduces surface tension,  gravity effects as well as normal viscous constraints. 
On the other hand,   the mean pressure  on interface $P'_m$   is obtained   up to a   constant  term, as the value of a continuous field on the surface. It is    related to  surface tension and gravity but  contains   inertial  and density effects as well (see below).   Using identities $P^{(1)'}=P'_m +[[P']]/2$ and $P^{(2)'}=P'_m -[[P']]/2$, equations \eqref{NStangentSources} can be re-formulated as
 \begin{equation}
\Sigma^{(1)}  = \frac{1}{\rho^{(1)} }   \frac{\partial P'_m }{\partial  s} +    \frac{1}{2 \rho^{(1)} }   \frac{\partial [[P']] }{\partial s}    
    +  \vec{t}^{1\to 2} \cdot  \frac{D\vec{u}}{Dt}
 \label{SigmaALternativeFinalNewBISphase1A} 
 \end{equation}
  \begin{equation}
\Sigma^{(2)}  =     - \frac{1}{\rho^{(2)} }   \frac{\partial P'_m }{\partial  s} +      \frac{1}{2 \rho^{(2)} }     \frac{\partial [[P']] }{\partial s}    
  -  \vec{t}^{1\to 2} \cdot  \frac{D\vec{u}}{Dt} 
 \label{SigmaALternativeFinalNewBISphase2A} 
 \end{equation}
Sources $\Sigma^{(1)}$ and $\Sigma^{(2)}$  depend  not only on  {\it  pressure gradients} along   the interface as for the total source $ \Sigma$ but  also on   a "tangential acceleration" term, that is  an {\it unsteadiness} of the  velocity at the interface.  This is similar to the generation of vorticity in a  boundary layer above  a solid wall.

  \vspace{0.9cm}

\noindent  It remains to express the mean pressure $ P'_m $   along the interface $(I)$ as an explicit  function of parameters $\sigma $, $g$, $ \mu_m$ and  $[[\mu]]$.  At a  given time $t$
and known velocity field, $\vec{u}(x,y,t)$, 
the pressure  field, $ P^{'} $, satisfies a Poisson equation  in each fluid 
 \begin{equation}
\upsilon \Delta P'   =-\partial_j(u_m\partial_m u_j)= - (\partial_j u_m) (\partial_m u_j)  
\label{PoissonPprime}
\end{equation}
and  two conditions  at the interface~:   the Young-Laplace law \eqref{MeqnrYoungLaplace2DNormalBIS} and  the continuity 
of  Lagrangian acceleration    across the interface \eqref{MeqnAcceleration}. More precisely the normal acceleration    across the interface \footnote{The continuity of    acceleration along the surface was already used   to get expression  
\eqref{MeqnIntegomegaJijVtot} for the source.} yields 
\begin{equation}
 [[\upsilon  n_i\partial_i   P^{'}  ]]=  [[  \mu  \upsilon n_i\partial_j \partial_j    u_{i} ]],
\label{derivatPcond}
\end{equation}
From  relation \eqref{diffpsurrho} with  $A=\mu  $  and $B=\upsilon u_{i} $, this equation can be re-expressed as 
\begin{equation}
[[\upsilon n_i\partial_i  P^{'}     ]]=      \frac{ [[\mu]]}{2} \bigg(  n_i \partial_j \partial_j  ( \upsilon^{(1)} u^{(1)}_{i})
 +   n_i\partial_j (\upsilon^{(2)}  \partial_j  u^{(2)}_{i})  \bigg) +  \mu_m [[ n_i\partial_j \partial_j   ( \upsilon u_i)]].
 \label{pressuregradientdsicontinous}
\end{equation}

\noindent  In the following we exhibit a  decomposition of  pressure  $ P^{'}$   into three fields
\begin{equation}
P^{'} \equiv \rho \Psi_{\rho_m} + P'_d +  P'_{c},
\label{decompPprime}
\end{equation}
Field  $\Psi_{\rho_m}$ and  its normal derivative   are continuous across the interface,  field $P'_d$ is 
discontinuous across the interface, $P'_{c} $  is continuous across the interface (but   its normal derivative is not). The  discontinuous field $P^{'}_d$  can be  itself    decomposed into   four discontinuous fields $ \Psi_{d \sigma}$,  $ \Psi_{d g} $,  $\Psi_{d \mu}$,  $\Psi_{d \rho}$ 
\begin{equation}
 P^{'}_{d} \equiv \sigma  \Psi_{d \sigma}  + g [[\rho]]   \Psi_{d g}   +  [[\rho]]     \Psi_{d \rho} + [[\mu]]   \Psi_{d \mu},
 \label{MeqPRESSUREDROPtris}
 \end{equation}  
Similarly, the    continuous  field $P'_c$  is    decomposed into a sum  
\begin{equation}
P^{'}_c \equiv  \sigma  \Psi_{\sigma}  + g [[\rho]]   \Psi_{g}  +  [[\rho]]     \Psi_{[[\rho]]} + [[\mu]]   \Psi_{[[\mu]]}   +   \mu_m  \Psi_{\mu_m}  
\label{Pprimecont}
\end{equation}  
of  the five continuous  fields   $\Psi_{\sigma}$,  $\Psi_{g} $,  $\Psi_{[[\rho]]}$  , $\Psi_{\mu_m}$, $\Psi_{ [[\mu]]}$   identified with   a different mechanism of vorticity production.

 \vspace{0.5cm}

\noindent  Let us now prove the above assertions. First we define  fields   $\Psi_{\rho_m}$,  $P'_d$  and $P'_{c}$ in \eqref{decompPprime}. The field  $\Psi_{\rho_m}$  satisfies  Poisson  equation
\begin{equation}
  \label{PoissonPprime}
  \Delta \Psi_{\rho_m}   = - (\partial_j u_m) (\partial_m u_j)   
\end{equation}
 in the {\it whole} fluid domain. Away from the interface $(I)$, the field  $\Psi_{\rho_m}$  satisfies the boundary condition of   $p^{(r)}/\rho^{(r)}-g \varphi$. This field and its normal derivative  are continuous across interface $(I)$. Such a solution exists and is unique    up to a meaningless constant  term, since  the r.h.s.   term in equation \eqref{PoissonPprime}  is only discontinuous  across interface $(I)$.  The field $P'_d$ satisfies a Laplace equation
\begin{equation}
  \label{eqpsic}
\Delta P'_d   = 0 
\end{equation}
and a  jump condition  obtained from equation \eqref{MeqnrYoungLaplace2DNormalBIS}
\begin{equation}
   [[P'_d]]  = - \sigma \kappa  +     g [[\rho]] \eta   + \,2\,[[\mu]] \bigg( \kappa \vec{u} \cdot \vec{n}^{1\to 2}
  - \, \frac{\partial }{\partial s} (\vec{u} \cdot \vec{t}^{1\to 2} )\bigg) - [[\rho]] \Psi_{\rho_m}.   
\end{equation}
To define $P^{'(1)}_d$ and $P^{'(2)}_d$ in each fluid we need one more relation at the interface. 
It is useful to impose $  P^{'(1)}_d +  P^{'(2)}_d = 0$. This latter condition and the  pressure  jump are   equally valid by imposing Dirichlet Boundary conditions such that
  \begin{equation}
 P^{'(1)}_d=-P^{'(2)}_d = \frac{1}{2} \left(- \sigma \kappa  +     g [[\rho]] \eta   + \,2\,[[\mu]] \bigg( \kappa \vec{u} \cdot \vec{n}^{1\to 2}
  - \, \frac{\partial }{\partial s} (\vec{u} \cdot \vec{t}^{1\to 2} )\bigg) - [[\rho]] \Psi_{\rho_m} \right) 
\end{equation}
 Finally,  the field $P^{'}_c$   satisfies  the Laplace equation
\begin{equation}
  \label{eqpsid}
  \Delta P'_{c}   = 0, 
\end{equation}
is   continuous across the interface while  its  normal derivative exhibits  a jump  
\begin{equation}
[[ \upsilon n_i\partial_i  P^{'}_c    ]]  =      \frac{ [[\mu]]}{2} \bigg(  n_i \partial_j \partial_j  ( \upsilon^{(1)} u^{(1)}_{i})
 +   n_i\partial_j (\upsilon^{(2)}  \partial_j  u^{(2)}_{i})  \bigg) +  \mu_m [[ n_i\partial_j \partial_j   ( \upsilon u_i)]] - [[  \upsilon n_i\partial_i  P^{'}_d ]]
 \end{equation}
 to comply with equation \eqref{pressuregradientdsicontinous}.  It is straightforward to check that the sum \eqref{decompPprime} satisfies the conditions of $P'$.  The discontinuous field $P^{'}_d$  
is   itself  the    sum  \eqref{MeqPRESSUREDROPtris}  of  four discontinuous fields $ \Psi_{d \sigma}$,  $ \Psi_{d g} $,  $\Psi_{d \mu}$,  $\Psi_{d \rho}$. Suppose each field  satisfies a Laplace  equation and a condition associated with a different mechanism of vorticity production namely defined    at any  interface points 
\begin{equation}
 \Psi^{(1)}_{d \sigma} =-  \frac{ \kappa}{2},~~~ \Psi^{(1)}_{d g} = \frac{\eta}{2},~~~~ \Psi^{(1)}_{d \mu} =  \bigg( \kappa \vec{u} \cdot \vec{n}
  - \, \vec{t} \cdot \vec{\nabla} (\vec{u} \cdot \vec{t})\bigg),~~~~~~ \Psi^{(1)}_{d \rho} =-\frac{1}{2}{\Psi_{\rho_m}},
  \label{LaplaceBCdiscontnew}
\end{equation}
and the exact opposite for $\Psi^{(2)}_{d \sigma} $, $ \Psi^{(2)}_{d g} $, $\Psi^{(2)}_{d \mu}$,  $\Psi^{(2)}_{d \rho}$.
The    fields $ \Psi_{d \sigma}$,  $ \Psi_{d g} $  only depend on the  geometry of the interface at time $t$. In addition to these parameters,  the field  $\Psi_{d \mu}$    is   also linearly dependent on   the velocity field  at time $t$, and the field $\Psi_{d \rho}$  on   the velocity field  at time $t$ but in a quadratic way instead.

 \vspace{0.5cm}
 
\noindent   Let us identify the five continuous  fields   $\Psi_{\sigma}$,  $\Psi_{g} $,  $\Psi_{[[\rho]]}$  , $\Psi_{\mu_m}$, $\Psi_{ [[\mu]]}$   in the sum \eqref{Pprimecont}   by  a different source of vorticity production.
They  all satisfy    Laplace  equation and two conditions  across the interface : continuity  $[[\Psi]]=0$ and 
\begin{equation}
[[\upsilon n_i\partial_i  \Psi_{\sigma}]]  =- [[\upsilon n_i\partial_i   \Psi_{d \sigma}]],~~~~~~~
[[\upsilon n_i\partial_i    \Psi_{g}]] =-  [[\upsilon n_i\partial_i     \Psi_{d g}]];
\label{cond1}
 \end{equation}
  \begin{equation}
[[ \upsilon n_i\partial_i  \Psi_{ [[\rho]]} ]]=- [[\upsilon n_i\partial_i    \Psi_{d \rho} ]],~~~~~~
[[\upsilon n_i\partial_i  \Psi_{ \mu_m}]]     =  [[  n_i\partial_j \partial_j  ( \upsilon u_i)]] 
\label{cond2}
 \end{equation}
\begin{equation}
[[\upsilon n_i\partial_i    \Psi_{ [[\mu]]}]]  =- [[ \upsilon n_i\partial_i   \Psi_{d \mu}]] 
+\frac{1}{2} \bigg( n_i \partial_j \partial_j  ( \upsilon^{(1)}  u^{(1)}_{i}) +   n_i\partial_j \partial_j  (\upsilon^{(2)}  u^{(2)}_{i})  \bigg) 
\label{cond3}
 \end{equation}
It is straigthforward to check that the sum \eqref{Pprimecont}  does satisfy the conditions of the continuous pressure field $P'_c$.  The  fields    $\Psi_{\sigma}$ and   $\Psi_{g} $   depend on the  geometry of the interface at time $t$ and on the density ratio. The  other  fields    depend  also on the velocity   at time~$t$~:  $\Psi_{[[\mu]]} $,  $\Psi_{\mu_m} $   depend linearly  on the amplitude of velocity,     $\Psi_{ [[\rho]]}$    on the square of this  amplitude.   The mean dynamic pressure  $P'_m=(P^{'(1)} +P^{'(2)} )/2 $  on an  interface $(I)$    is thereafter  obtained as 
 \begin{equation}
P'_m = \rho_m   \Psi_{\rho_m} +  P'_{c}, 
\label{bi}
\end{equation}
Details  on numerical points are included in appendix~\ref{AppendixNumericalimplementation}. 

 \vspace{0.2cm}

 \section{ Vorticity production: the full expression}
 \label{vorticityproductionexpression}
 
\noindent The vorticity production  $\Sigma$   can    be expressed  using  equation  \eqref{SigmaALternativeNew}
  \begin{equation}
   \Sigma  =  \frac{\partial \Psi_{\Sigma}}{\partial s},~~~    \Psi_{\Sigma} \equiv  [[\upsilon]] \rho_m      \Psi_{\rho_m} +  [[\upsilon]]  P'_{c}   + \upsilon_m  [[P']]. 
  \label{SigmaALternativeNewBISIS} 
 \end{equation}
Combining  equations   \eqref{MeqnrYoungLaplace2DNormalBIS}, \eqref{Pprimecont} and \eqref{SigmaALternativeNewBISIS},    this quantity may be  written as  an explicit  function  of parameters $\upsilon_m$, $\sigma $, $g$, $ \mu_m$,  $[[\mu]]$
  \begin{equation}
   {\Sigma} = \frac{\partial \Psi_{\Sigma} }{\partial s},~~~\Psi_{\Sigma}=- \sigma \upsilon_m  \kappa    + [[\upsilon]] \bigg( L_{\upsilon}  +   N_{\upsilon}\bigg) +  [[\mu]]  \bigg( L_{[[\mu]]}  +  N_{[[\mu]]} \bigg)
   \label{MeqnSOmegaInterfaceNEW}
\end{equation}
with $L_{\upsilon}$ and $L_{[[\mu]]}$ containing terms linear  with  the  perturbation amplitude
 \begin{equation} 
L_{\upsilon}  \equiv       \sigma    \Psi_{\sigma}     
+ g [[\rho]]     \Psi_{g}      
+   \mu_m    \Psi_{\mu_m}   
 -     \rho_m g     \eta,~~~~~        
  L_{[[\mu]]}  \equiv    [[\upsilon]]     \Psi_{[[\mu]]}     
 - \,2  \upsilon_m\,  \frac{\partial  (\vec{u} \cdot \vec{t}^{1\to 2} )}{\partial s}       
 \label{MeqnSOmegaInterface2}
\end{equation}
and $N_{\upsilon}$, $N_{[[\mu]]}$ containing  non-linear  terms with respect to   perturbation amplitude
\begin{equation} 
N_{\upsilon} \equiv  \rho_m      \Psi_{\rho_m}  +  [[\rho]]      \Psi_{[[\rho]]} 
       ,~~~~~~N_{[[\mu]]} \equiv   \,2  \upsilon_m\,      \kappa \vec{u} \cdot \vec{n}^{1\to 2}    \label{MeqnSOmegaInterface3bis}
\end{equation}
Each  term in equations   \eqref{MeqnSOmegaInterfaceNEW}  \eqref{MeqnSOmegaInterface2} \eqref{MeqnSOmegaInterface3bis} reveals  the importance of  surface tension, viscous stresses, gradient forces such as  gravity,  and  inertial forces     on the production of vorticity across the interface.  Vorticity production is  sum  of these different mechanisms valid at each given time. The   accumulated effect over time of these different mechanisms {\it viz}  the produced vorticity   is  obviously not  linear.  We compute  the sources on the interface   finally
reducing the problem to  the evaluation of $\kappa$, $\eta$ and the various functions $\Psi$ at the domain boundaries.
This method shares some features of boundary integral method  adapted to high Reynolds number situations.  In  boundary integral methods,  the sheet strength along the interface  is first computed through a Fredholm integral equation obtained  through  the jump of the normal component of stress and  the tangential  Navier-Stokes component along the surface. Second the vorticity is diffused  near the interface to ensure continuity of tangential shear.  In the present case,  there is  indeed a boundary layer   but not a vortex sheet although we still perform a two-step process: First we  compute the circulation along the surface which is related to the normal stress. This provides on the average vorticity  on the surface $\omega_m\equiv (\omega^{(1)}+ \omega^{(2)})/2$; Second we get the vorticity on each side by using the conditions related to continuity of shear 
\begin{equation}  
[[ \mu {\omega}^{(\pi)}_l]]=  [[2\mu]] {W}^{(\pi)}_l 
~\hbox{with}~~{\omega}^{(\pi)}_l \equiv  {\omega}_l - ({\omega}_k {n}_k) {n}_l,~~~~
W^{(\pi)}_l \equiv - \epsilon_{lpq}  n_p n_k \partial_q u_k.
\label{MeqnSUrfacenormalvectorevolutionbis4}
\end{equation}

 \vspace{0.5cm}

\noindent  One   could  introduce the above ideas   inside  a Lagrangian numerical  scheme based on vorticity and  pressure  fields that simulates two fluids  separated by a     sharp interface.   For this purpose, it is  not  necessary  to decompose the expression as   above and one     numerically solves one   Poisson equation and    four  Laplace equations. First  the  Poisson equation \eqref{PoissonPprime} for $\Psi_{ \rho_m}$ over the whole space; second  a  Laplace equation  in each fluid domain  for  the field $\Psi_{d}$ with Dirichlet conditions  
\begin{equation}
 \Psi^{(1)}_{d } =-  \sigma \frac{ \kappa}{2} + g [[\rho]] \frac{\eta}{2} -\frac{[[\rho]]}{2}{\Psi_{\rho_m}} + [[\mu]]\bigg( \kappa \vec{u} \cdot \vec{n}  - \, \vec{t} \cdot \vec{\nabla} (\vec{u} \cdot \vec{t})\bigg),~~~~~ \Psi^{(2)}_{d }= -\Psi^{(1)}_{d }. 
  \label{LaplaceBCdiscontnewbis}
\end{equation} 
Third a Laplace equation  for  the field   $\Psi_{\mu_m}$ over the whole space  with conditions   \eqref{cond2} and fourth the  Laplace equation   over the whole space for the field  $\Psi_{c}   $   with condition 
 \begin{equation}
[[n_i\partial_i  (\upsilon  \Psi_{c})]]    = - [[ n_i\partial_i  (\upsilon  \Psi_{d})]] + \frac{1}{2} \bigg(  \upsilon^{(1)} n_i \partial_j \partial_j  u^{(1)}_{i} + \upsilon^{(2)}   n_i\partial_j \partial_j  u^{(2)}_{i}  \bigg). 
 \end{equation}
 These computations could  be also used  in  Eulerian numerical methods to provide a better approximation of   velocity gradients .

 \vspace{0.5cm}

\noindent  The   equations for vorticity production  could be alternatively put  in dimensionless form using the  characteristic dimensional    velocity  $U_0$,  length  $L_0$  together with  the average density $\rho_m =(\rho^{(1)}+\rho^{(2)})/2 $.   Furthermore the various source terms are non-dimensionalized as follows 
\begin{equation}
\Psi_{\Sigma}=  U^2_0  {\hat \Psi}_{\Sigma};~~ \Psi_{\sigma}= \frac{1}{L_0} {\hat \Psi}_{\sigma};
~~ \Psi_{g}=  L_0  {\hat \Psi}_{g};~~~\Psi_{\rho_m}= U^2_0 {\hat \Psi}_{\rho_m},~~\Psi_{[[\rho]]}= U^2_0 {\hat \Psi}_{[[\rho]]},~~\Psi_{\mu_m}=  \frac{U_0}{L_0} {\hat \Psi}_{\mu_m};
\label{Meqn1:999Nondimensionalization12}
\end{equation} 
Note that  $\Psi_{x}$ possesses similar dimension than $\Psi_{d x}$  for $x=\sigma,g, \rho$;  $\Psi_{\mu_m}$ and  $\Psi_{[[\mu]]}$ similar dimensions than  $\Psi_{d \mu}$ and $\Psi_{[[\rho]]}$ similar dimensions than $\Psi_{\rho_m}$.   The  dimensionless production source reads as   
$\Sigma  =  \frac{\partial \Psi_{\rm{inv,L}}}{\partial s}
+\frac{\partial \Psi_{\rm{inv,NL}}}{\partial s}
+\frac{\partial \Psi_{\rm{visc}}}{\partial s}
$  
 with      
\begin{eqnarray}
  \hat \Psi_{\rm{inv,L}} &=& - \frac{1}{1 - A_{tw}^2} \frac{ \kappa}{We}     
+   \frac{2 A_{tw}}{1 - A_{tw}^2}  \frac{1}{We}   \hat \Psi_{\sigma}      
-  \frac{4 A^2_{tw}}{1 - A_{tw}^2} Fr  \hat  \Psi_{g} 
-  \frac{2 A_{tw}}{1 - A_{tw}^2} Fr  ~  \hat \eta,
   \label{Sigmasource1} \\
  \hat \Psi_{\rm{inv,NL}} &=&   
 \frac{2 A_{tw}}{1 - A_{tw}^2}  \hat \Psi_{\rho_m}
-  \frac{4 A^2_{tw}}{1 - A_{tw}^2} \hat  \Psi_{[[\rho]]},
   \label{Sigmasource3}\\
   \hat \Psi_{\rm{visc}} &=&   \frac{[[\upsilon]] \mu_m}{\nu_m}  \frac{1}{Re} \hat  \Psi_{\mu_m} 
  + \frac{[[\upsilon]]~[[\mu]]}{\nu_m}     \frac{1}{Re} \hat   \Psi_{[[\mu]]}  
+  \frac{\upsilon_m [[\mu]]}{\nu_m}  \frac{2}{Re} \bigg( \kappa  \hat \vec{u} \cdot \vec{n}^{1\to 2} - \frac{\partial (\hat \vec{u} \cdot \vec{t}^{1\to 2} )}{\partial s}  \bigg).
   \label{Sigmasource2} \nonumber \\
\end{eqnarray}
 Several dimensionless number  appear: the  Reynolds,  Weber number and  Froude numbers as well as the Atwood number\footnote{Note  we used  the equalities $\rho_m  \upsilon_m =  \frac{1}{1 - A^2_{tw}}$ ,  $[[\upsilon]]\rho_m=\frac{2 A_{tw}}{1 - A^2_{tw}}$, and $ [[\rho]][[\upsilon]]=-\frac{4 A^2_{tw}}{1 - A^2_{tw}}$. }
\begin{equation}
   Re \equiv \frac{  U_0 L_0}{\nu_m}, ~~We \equiv  \frac{ \rho_m U_0^2 L_0 }{\sigma},~~Fr \equiv  \frac{ gL_0}{U_0^2}, 
   ~~~A_{tw} \equiv   \frac{\rho^{(2)}-\rho^{(1)} }{\rho^{(2)}+\rho^{(1)}}
\label{MeqnLambOseenReyWe}
\end{equation}

 \subsection{ Asymptotic case $A_{tw} \to 1$.}
 \label{vorticityproductionextremecaserho1rho2}
 
  \noindent  When  if $\nu^{(2)}/\nu^{(1)} =O(1)$ and one fluid is much lighter  than the other one e.g.  fluid 1 is much lighter than fluid 2 $ {\rho^{(2)} }>> \rho^{(1)}$,   the situation is simpler from a mathematical viewpoint since  a small parameter  $ \epsilon \equiv \frac{ \upsilon^{(2)}}{ \upsilon^{(1)}}=\frac{ \rho^{(1)}}{ \rho^{(2)}} <<1$ exists.  This case is adequate  for the  air-water interface since $\rho^{(1)} / \rho^{(2)} \approx  10^{-3}$,
and  $\nu^{(1)}/\nu^{(2)} \approx 10^{-1}$.  Note that  $\epsilon \equiv \frac{1-A_{tw}}{1+A_{tw}}$.

 \vspace{0.5cm} 
  
\noindent   In appendix \ref{vorticityproductionextremecaserho1rho2}, the full computations  show that
 for $A_{tw} \to 1$ and $Re>>1$, one obtains 
 \begin{equation}
\Psi_{\Sigma} \approx   - \frac{1}{2} \frac{ \kappa}{We}  -   Fr~\hat \eta +    \hat \Psi_{\rho_m}~~~\hbox{for}~~~Re>>1
   \label{Sigmasource2nodimtot1}
   \end{equation}

   Reversely  for Stokes hydrodynamics   $Re<<1$, the viscous term   $\Psi_{\rm{visc}}$ is the leading contribution  
  \begin{equation}
\Psi_{\Sigma} \approx   \frac{[[\upsilon]] \mu_m}{\nu_m}  \frac{1}{Re}    \Psi_{\mu_m} 
  + \frac{[[\upsilon]]~[[\mu]]}{\nu_m}     \frac{1}{Re}    \Psi_{[[\mu]]}  
+  \frac{\upsilon_m [[\mu]]}{\nu_m}  \frac{2}{Re} \bigg( \kappa  \hat \vec{u} \cdot \vec{n}^{1\to 2} - \frac{\partial (  \vec{u} \cdot \vec{t}^{1\to 2} )}{\partial s}  \bigg)
   \label{Sigmasource2nondim}
\end{equation}

 \vspace{0.5cm}

 \subsection{Case  $A_{tw} \to 0$.}

\noindent  When the density of both fluid is equal the equations above can be further simplified to obtain $\Psi_{\Sigma}$ as       
\begin{equation}
\hat \Psi = -  \frac{ \kappa}{We}+ \frac{\upsilon_m [[\mu]]}{\nu_m}  \frac{2}{Re} \bigg( \kappa  \hat \vec{u} \cdot \vec{n}^{1\to 2} - \frac{\partial (\hat \vec{u} \cdot \vec{t}^{1\to 2} )}{\partial s}  \bigg)
\end{equation} 
In the case of $[[\mu]]=0$ the only source of vorticity is due to curvature changes on the interface. 
Steady state solutions with $\Psi_{\Sigma} = 0$ and non-uniform values of $\kappa$ are admitted if $\mu^{(1)} \ne \mu^{(2)}$ and $u\ne 0$.

\section{An analytical example: Viscous gravito-capillary flows.   }
 \label{GravitocapillaryAnaly}

 \noindent  In order to show the possible applications of the above decomposition,  let us   consider  various  examples  of   two   immiscible fluids  separated by an interface $(I)$  located at $ y=\eta(x,t)$, the lighter fluid $1$ being  above the heavier  fluid $2$.     Gravity $\vec{g}= -g \vec{e}_y$   (variable $y$ denoting the upward vertical position) is present together or without  surface  capillary forces.    In the examples studied,  either  vorticity sources can be computed exactly to get quantitative predictions or else can be qualitatively evaluated    providing an understanding of    the  observed dynamics.   The   example  presented in this section is   in the linear frame and explicit computations can be performed based on the  source terms leading to   the dispersion relation for viscous gravito-capillary waves. To get this result, we could use the continuity of velocity and jump in stress tensor. We could use instead the continuity of velocity and jump of vorticity and the increase of vorticity.

\vspace{0.5cm}

    \noindent Consider an infinitesimal amplitude wave. The  interface is   perturbed  at  $y =\eta(x,t)$ from its flat equilibrium position. Such a  wave  can be  decomposed in  Fourier modes $ y =\eta(x,t) = B_0 a_0   \exp{i (kx-\varpi(k) t)}$ where   $k$  is  a real wavenumber and $\varpi(k)$  a complex pulsation and a wave slope $B_0 a_0 k <<1 $.   Gravity  force   still  generates   vorticity on the  interface which thereafter    diffuses or is advected  into the bulk producing a net motion.  Owing to small amplitude,  we set \footnote{The curvilinear coordinate $s$   is   defined on $(I)$  so  that  it  increases in the direction of $\vec{t}^{1\to 2}$ which is axis along $x$ in the present case.}  $x=s$ in  the source \eqref{MeqnSOmegaInterfaceNEW}  and retain  only the linear terms    with respect to the amplitude perturbation. 
Curvature $ \kappa$   is  positive   when the  centre  of curvature lies in   phase 2 which  is   located  at $y \le \eta(x,t)$. For linear case, this implies $\kappa=- \frac{\partial^2 \eta}{\partial x^2}$ that is $\kappa=k^2 \eta$.  This yields the  expression for the source 
 \begin{equation}
   {\Sigma} =  ik  \Psi_{\Sigma}  ,~~~\Psi_{\Sigma}=    -\sigma \upsilon_m k^2   \eta   
 +    [[\upsilon]]         L_{\upsilon}    +  [[\mu]]   L_{[[\mu]]}      
 \label{MeqnSOmegaInterfaceLINEARVISCOUS}
\end{equation} 
 \begin{equation} 
L_{\upsilon}  \equiv      \sigma    \Psi_{\sigma}     
 -     \rho_m g     \eta + g [[\rho]]     \Psi_{g}      
+   \mu_m    \Psi_{\mu_m}, ~~~~~
  L_{[[\mu]]}  \equiv   [[\upsilon]]     \Psi_{[[\mu]]}     
 - \,2  \upsilon_m\,  \ ik {u}_x. 
 \label{MeqnSOmegaInterface3}
\end{equation}
To compute  explicitly  $ \Psi_{\sigma}$, $\Psi_{g}$, $\Psi_{\mu_m}$, $\Psi_{[[\mu]]}$ and ${u}_x$, 
it is necessary to express  the velocity field  as a function of  the interface motion.

\vspace{0.5cm}

 \noindent    First let us compute  the velocity field for a  viscous capillary gravity wave   characterized by an interface located at  $ y =\eta(x,t)$.
First for small amplitudes, the linearization in the two phases yields 
\begin{equation}
 \rho^{(r)}  i\varpi {u}_i^{(r)}  =  \partial_j  P^{('r)}   - \mu^{(r)}\Delta u^{(r)}_i,~~~~~\partial_i u^{(r)}_i=0,~~~~r=1,2. 
\label{Meqn1:999Linear}
\end{equation}
By taking the divergence of the above expression, it is seen  that pressure $P'$ is harmonic. It is thus possible to find  a harmonic  function $\phi$ such that 
\begin{equation}
 \rho^{(r)}  i\varpi \phi^{(r)}    =  P^{('r)},~~~~r=1,2. 
\label{MeqnHELMHOLTZDECOMPPhi}
\end{equation}
In addition,  incompressibility condition leads to the existence of a streamfunction $\psi^{(r)}$ such that
\begin{equation}
 {u}^{(r)}_x={ \partial_x}\phi^{(r)} - { \partial_y}\psi^{(r)};~~~ {u}^{(r)}_y={ \partial_y}\phi^{(r)} +{ \partial_x}\psi^{(r)},~~~~r=1,2.
\label{MeqnHELMHOLTZDECOMP}
\end{equation}
This is nothing else  but a simplified Helmholtz decomposition \citep{wu1995theory}.  Vorticity is equal to 
$\omega^{(r)}= \Delta {\psi}^{(r)} $ and  the Navier-Stokes equation simply becomes
 \begin{equation}
- i\varpi \psi^{(r)} =     \nu^{(r)} \Delta   {\psi}^{(r)} +C(t)= \nu^{(r)} \omega^{(r)} +C(t),
\label{MeqPSilinearMAIN}
  \end{equation}
where $C(t)$ is a function of time.  Classically  this constant is set to zero by a simple redefinition  $\psi^{(r)} \to \psi^{(r)} + \int^{t}_0 C(t') dt'$.  The potential field being   harmonic, its dependency  on $y$ is determined. Similarly the streamfunction $\psi$ is governed by \eqref{MeqPSilinearMAIN}    possesses a  clear $y$-dependency 
   \begin{equation}
  \phi = 
  \begin{cases}
    A^{(1)}\exp{ \bigg( i (kx-\varpi t) -|k| y \bigg)} & \\
    A^{(2)} \exp{ \bigg( i (kx-\varpi t)  + |k| y \bigg)},&  
  \end{cases}
  ~~~\psi = 
  \begin{cases}
    B^{(1)}\exp { \bigg(i (kx-\varpi t)   - \kappa^{(1)} y \bigg)}
    & \text{if  $0 < y$}\\
     B^{(2)}\exp { \bigg(i (kx-\varpi t) +  \kappa^{(2)} y\bigg)}
     & \text{if $y\le 0$}
   \end{cases}
    \label{MeqModenormalFormphiappen}
  \end{equation}
  where quantity $\kappa^{(r)}$ is  a complex number  with a real positive part such that
  \begin{equation}
 [\kappa^{(r)}]^2 = k^2 - i \frac{ \varpi(k)}{\nu^{(r)}},~~~~r=1,2. 
  \label{MeqModenormalFormpskappaappen}
  \end{equation}
In the following we use the notations
\begin{equation}
  b^{(r)} \equiv k -   \frac{k}{|k|}\kappa^{(r)},~~~~r=1,2. 
  \label{MeqModenormalFormpdefinitionb}
  \end{equation}

\vspace{0.5cm}

\noindent   The continuity of  velocity field,  the jump of vorticity   across the interface as well as  the kinematic condition on the interface     yield   
coefficients $|k| A^{(1)}$,  $|k| A^{(2)}$, $B^{(1)}$,  $B^{(2)}$ as functions of amplitude $\eta_0$. 
Finally  the source $\Sigma = ik  \Psi_{\Sigma}  \exp { i (kx-\varpi(k) t) }$   in  \eqref{MeqnSOmegaInterfaceLINEARVISCOUS}    is expressed 
as a function  of $|k|A^{(1)}$,  $|k|A^{(2)}$ and $\eta_0$ {\it via} the conditions at $y=0$ (see appendix~\ref{Viscoussourcefieldgravitywave} for computations)  
\begin{equation}
 {|k|} \Psi_{\Sigma} = \big(-2{\varpi^2_{inv}}   +\frac{[[\mu]]}{\rho_m}  i (\kappa^{(1)}-\kappa^{(2)}) \varpi \big) \eta_0  +  \alpha_1 {|k|} A^{(1)}  + \alpha_2 {|k|} A^{(2)}
  \label{MeqnAppendixSOmegaInterfaceVISCOUSsimplifiedappend}
\end{equation} 
where 
 \begin{equation}
 \alpha_1 = i A_{tw} \varpi   +\frac{ [[\mu]]}{\rho_m} k b^{(1)},~~~~~
  \alpha_2 =   i A_{tw} \varpi    +\frac{[[\mu]]}{\rho_m}  k b^{(2)}       
 \label{MeqnAppendixSOmegaInterfaceVISCOUSDEFINITIONALPHA}
\end{equation} 
and   the inviscid pulsation $ \varpi_{inv}$ 
\begin{equation}
 \varpi^2_{inv}(k)    \equiv  \bigg( A_{tw}  g +\frac{\sigma k^2}{\rho^{(1)} +\rho^{(2)}} \bigg){\mid k\mid}.
 \label{eqndisperrelaAnnexe}
\end{equation}

\vspace{0.5cm}

\noindent  Each Fourier component  evolves independently and equation \eqref{MeqnIntegomegaJijVtotbis} gets simplified for a unique Fourier component   
   \begin{equation}
-i \varpi  \left(\int_{A}\omega\,\,\text{d}x  \text{d}y\right)
  =     -\int_{C} J_{j}\,n_{j}\,\text{d}y_c
   +\int_{I} \Sigma \,\text{d}x,
   \label{MeqnIntegomegaJijVtotbisLinearized}
 \end{equation}
 where the loop $(C)$ lies along the $y$-axis at  $x$ and $x+dx$ and is closed at   $\pm \infty$ in $y$, path $(I)$ corresponds to  a stretch   $ds=dx$  of the interface. The l.h.s. integral and   first r.h.s. integral  can be expressed  as  line integrals  
  \begin{equation}
\gamma^{(2)} \equiv \int^{0}_{-\infty}\omega^{(2)} \, \text{d}y,~~~
\gamma^{(1)} \equiv \int^{\infty}_{0}\omega^{(1)} \, \text{d}y, 
 \label{MeqnIntegomegaJijVtotbis11}
 \end{equation}
so that equation \eqref{MeqnIntegomegaJijVtotbisLinearized} becomes 
   \begin{equation}
(i \varpi - \nu^{(1)}k^2) \gamma^{(1)}  + (i \varpi -  \nu^{(2)} k^2)\gamma^{(2)}+\Sigma=0
   \label{MeqnIntegomegaJijVtotbisLinearizedBIS}
 \end{equation}
 The first r.h.s. term is clearly due to the decay of circulation by diffusion through the outer boundaries. 
  Since $\omega^{(r)}=  \Delta \psi^{(r)}=-(k^2-(\kappa^{(r)})^2)\psi^{(r)}$, the integration leads to     
\begin{equation}
\gamma^{(1)}=  - \frac{(k^2-(\kappa^{(1)})^2)}{\kappa^{(1)}} B^{(1)};
 ~~~\gamma^{(2)}= -  \frac{(k^2-(\kappa^{(2)})^2)}{\kappa^{(2)}}   B^{(2)};
\label{circulationdefviscouslinear1}
\end{equation}
 Finally noting that
  $$
  (i \varpi - \nu^{(r)}k^2) \frac{(k^2-(\kappa^{(r)})^2)}{\kappa^{(r)}}=-i \varpi \kappa^{(r)}
 $$
  the dynamics reduces to  
 $$
i \varpi \bigg( \kappa^{(1)}  B^{(1)}  + \kappa^{(2)}  B^{(2)}\bigg)+\Sigma=0
 $$
or using \eqref{eq:Vinterface2BIS}  
 \begin{equation}
  i \varpi \bigg( |k| A^{(2)}- |k|A^{(1)}  \bigg)+|k| \Psi_{\Sigma} =0, 
      \label{MeqnIntegomegaJijVtotbisLinearizedBIS3}
 \end{equation} 
which once combined with   \eqref{MeqnAppendixSOmegaInterfaceVISCOUSsimplifiedappend}  yields  
\begin{equation}
[i \varpi \beta_1 + \beta_2] |k| A^{(1)}
+ [ i \varpi \beta_3 +\beta_4]  |k| A^{(2)}
+[  -2{\varpi^2_{inv}}    + i \varpi \beta_5  ] \eta_0  
 =0
  \label{MeqnAppendixSOmegaInterfaceVISCOUSsimplifiedappenNEW}
\end{equation}  
$$
\beta_1= -1+ A_{tw},~~~\beta_2=\frac{ [[\mu]]}{\rho_m} k b^{(1)},~~~~
\beta_3= 1+ A_{tw},~~~\beta_4=  \frac{[[\mu]]}{\rho_m}  k b^{(2)},~~~~
\beta_5= \frac{[[\mu]]}{\rho_m}  k (b^{(2)}-b^{(1)}).
$$  
From equation \eqref{MeqnAppendixSOmegaInterfaceVISCOUSsimplifiedappenNEW},   \eqref{MeqnNormalParallelvorticitebislinea2} and 
\eqref{MeqnNormalParallelvorticitebislinea2BIS},    it is tedious but straightforward to obtain the 
well-known dispersion relation for gravito-capillary    waves\citep{prosperetti1981motion}.
\begin{equation}
-\varpi^2   +{\varpi^2_{inv}} +  \frac{C_2\varpi^2 +i \varpi C_1+C_0 }{(\rho^{(2)}b^{(1)} +\rho^{(1)} b^{(2)}) }=0
  \label{MeqnAppendixSOmegaInterfaceVISCOUSNEW1}
\end{equation}  
$$
C_0=  2 k^3\frac{[[\mu]]^2 }{\rho_m}b^{(1)} b^{(2)},~~~~C_2= 2k \frac{\rho^{(1)}\rho^{(2)}}{\rho_m},~~~~
C_1= 2k^2\frac{[[\mu]]^2}{\rho_m} (\rho^{(2)}b^{(1)} -\rho^{(1)} b^{(2)}).
$$

\section{ Gravity Waves : numerical  non-linear   cases.   }
\label{bumpflowgrav}

\noindent   The    flow examples  proposed in this section are gravity waves without surface tension but and  in contrast to the previous section,   they   are  typically in a nonlinear regime. In that case  source terms lead to qualitative understanding or  constitutes a test for numerical simulations. Initially  the fluid is   at rest, and the interface     
\begin{equation}
y=\eta(x,t=0)  =    B_0{a_0} \exp(-(x/a_0)^2)
\label{Meq1appendix}
\end{equation}
is disturbed by a large initial  amplitude $B_0 a_0$ (here   $|B_0| = 2.5$  or $|B_0|=5$) and it is periodic along $x$ of period  $La_0$, $L$ being large enough in practice $L=5|B_0|$.    The flow is computed  by solving Navier--Stokes equations    { \it via} the two-phase approach based on Volume of Fluid \cite{tryggvason2011direct}.
More specifically,  we use the code Basilisk \citep{popinet2018numerical}  inside a  two-dimensional domain 
$(x,y) \in [-L a_0/2,L a_0/2 ]\times [-L a_0/2,L a_0/2 ]$  with  a regular grid  of size $\Delta x = 0.01 a_0$.  Periodic boundary conditions are imposed  at the  right/left side  and impenetrability and slip wall conditions ($v=0$ and $\partial_y u=0$)  at the top and bottom side.  In order to simplify this multi-parameter situation,   dynamical viscosity is assumed    identical in both fluids : velocity, vorticity and  tangential   stress  are thus  continuous across the interface but a vorticity  flux  \eqref{MeqnSOmegaInterfaceNEW}  is nonetheless present.

\vspace{0.2cm}

\noindent  For  each flow,  a characteristic   length is given by    size $a_0$,  a  characteristic  velocity given by   $U_0 \equiv   a_0  \sqrt{A_{tw} g {\pi}/{a_0}} $       the  product of $a_0$ by the inviscid   pulsation~ $\varpi =\varpi_{inv}(k)$ at $k = {\pi}/{a_0}$ and the average density ${\rho_m} $. Based on such   dimensional  equations,   
the dynamics written in  dimensionless quantities is governed by a wave slope $B_0$, a Reynolds number and  a  density ratio or an Atwood number.
 The simplified vorticity flux reads in dimensionless form as 
  \begin{equation}   
 \Sigma  =  \frac{\partial \hat \Psi_{\Sigma}}{\partial s},~~~~~\hat  {\Psi}_{\Sigma} =     \frac{2}{1 - A^2_{tw}} (  \hat  L_{\upsilon} + \hat N_{\upsilon}) 
     \label{MeqnSOmegaInterfaceNEWnondimensional}
\end{equation}
with   $\hat L_{\upsilon}$  terms linear   with respect to the  perturbation amplitude
 \begin{equation} 
   \hat L_{\upsilon} \equiv \frac{A_{tw}}{Re}  \hat  \Psi_{\mu_m} - \frac{1}{\pi} \hat \eta  - \frac{2 A_{tw}}{\pi} \hat  \Psi_{g}         
\label{MeqnSOmegaInterface2NONDIM}
\end{equation}
and $N_{\upsilon}$    non-linear  terms with respect to   perturbation amplitude
\begin{equation} 
  \hat N_{\upsilon} \equiv   A_{tw} (   \hat \Psi_{\rho_m}  -   2  A_{tw}   \hat  \Psi_{[[\rho]]} ).
  \label{MeqnSOmegaInterface3bisNONDIM}
\end{equation}
In what follows,  numerical simulations are presented for  density  ratio $r_{\rho}=2$  and $r_{\rho}=10 $ or respectively  Atwood numbers $A_{tw} =1/3$   and $A_{tw} =9/11$.

 \subsection{Initial time evolution : quantitative predictions}

\vspace{0.2cm}

   \noindent First  let us examine the circulation    for $x \in [0,L/2]$  and near $t=0$.  During that period,  the fluid is almost at rest and  vorticity is zero initially. As a consequence equations \eqref{MeqnIntegomegaJijV1}  read
 \begin{equation}
\frac{\partial}{\partial t}\left(\int_{[0,L/2]}\omega\,\text{d}x  \text{d}y\right)
  =    -\int_{(C) } J_{j}\,n_{j}\,\text{d}s_c + \int_{(I) }  \Sigma \,\text{d}s,~~~r=1,2
  \label{MeqnIntegomegaJijV1linearappend}
\end{equation}
where   $(I)$  denotes   the interface for  $0\le x \le L/2$  and   $(C) $  a loop around the  positive part $0\le x \le L/2$.  It is easy to show that the first r.h.s. is zero so that
 \begin{equation}
 \Gamma_{[0,L/2]}(t)    \equiv   \int_{[0,L/2]}   \omega (x,y,t) dxdy, 
\end{equation}
evolves according to 
\begin{equation}
  \Gamma_{[0,L/2]}(t)    = C ~t~~~\hbox{with}~~~~C  \equiv   \int_{(I)}   {\Sigma}_A(s)\, ds 
 \end{equation}
$ \hat  {\Sigma}_A$ denoting   the  source at $t=0$.  Initially  the fluid is at rest (this  configuration is   denoted below  as configuration A)  and  $ \hat  {\Sigma}_A$  takes  the simple expression 
\begin{equation}
  \hat \Sigma_A  =  \frac{\partial \hat \Psi_{\Sigma}}{\partial s},~~~~~\hat  {\Psi}_{\Sigma} =     -\frac{2}{\pi}\frac{1}{1 - A^2_{tw}} ( \hat \eta  +2 A_{tw} \hat  \Psi_{g} )
  \label{SigmaAfull}
\end{equation}
and since $ \hat  \eta(x=0) - \hat  \eta(x=L/2) \approx  B_0 $, coefficient $C$  is equal to 
 \begin{equation}
   C(r_{\rho}) \approx  \frac{2}{\pi} \frac{1}{1 - A^2_{tw}}  ( B_0 + 2 A_{tw} \Delta \hat \Psi_g),~~~\Delta \hat \Psi_g \equiv  \hat \Psi_g(x=0) - \hat \Psi_g(x=L/2) 
     \label{MeqnSOmegaInterfaceNEWnondimensionalt0ineteg}
\end{equation}
Finally note that the linearized  expression of  \eqref{SigmaAfull} yields (to be used later) $\hat \Psi_{\Sigma} = -\frac{2}{\pi} \hat \eta$.

 \vspace{0.5cm}

\begin{figure}
   \begin{center}
\includegraphics[width=0.4 \textwidth]{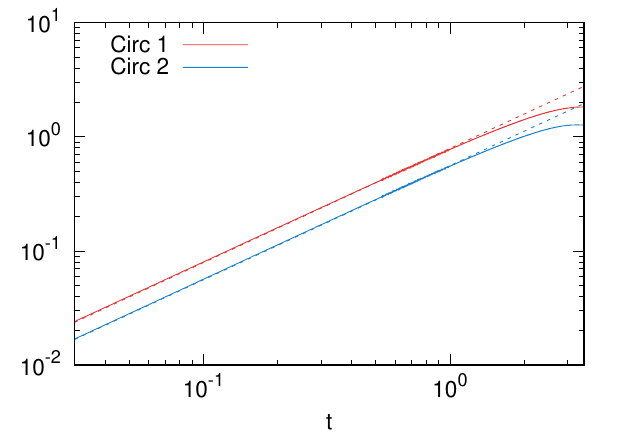}
\includegraphics[width=0.4 \textwidth]{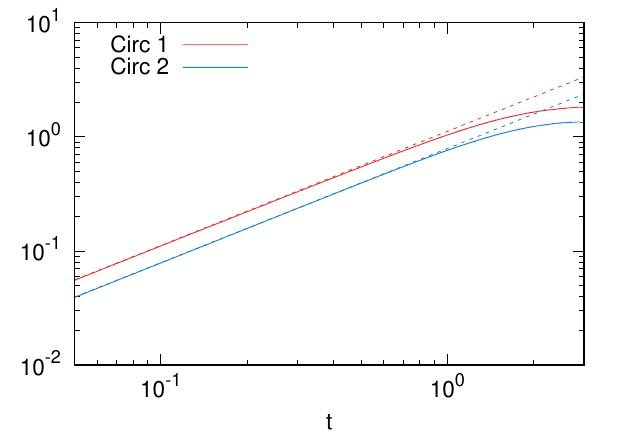}
\end{center}
\caption{ Nonlinear   gravity perturbation characterized by  $A_{tw}=1/3$ ,  $Re=10^3$  with (left) $B_0=2.5$  and (right) $B_0=-2.5$ : Circulations $ \Gamma^{(1)}_{x \ge 0}(t)$, $ \Gamma^{(2)}_{x \ge 0}(t)$  as a function of time.
Numerical values from DNS simulations are displayed using solid lines, and  theoretical values \eqref{MeqnIntegomegaJijVtotbisLinearizedBISfiniteampliTRISIS} 
 by dashed lines. }
\label{figCircB25}
 \end{figure}

\begin{figure}
   \begin{center}
\includegraphics[width=0.4 \textwidth]{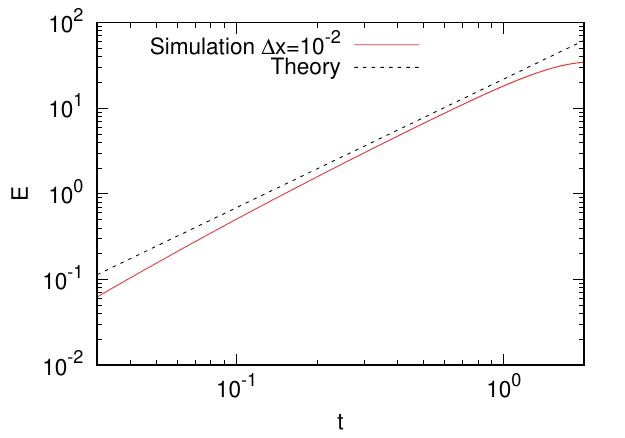}
\includegraphics[width=0.4 \textwidth]{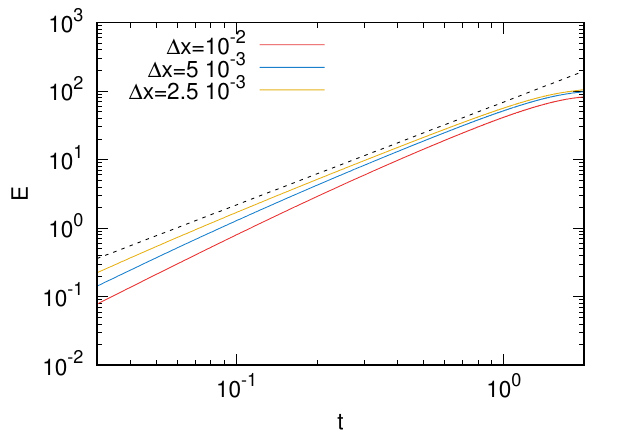}
\end{center}
\caption{Nonlinear   gravity perturbation characterized by  $A_{tw}=1/3$ and $B_0=-2.5$ with  (left) $Re=10^3$ and (right) $Re=10^4$: 
Temporal evolution of   the  total enstrophy $E$.  solid lines are   DNS simulations  values and  dashed lines ares theoretical values   \eqref{MeqnIntegomegaJijVtotbisLinearizedBISfiniteampliTRISIS}. }
\label{figEnsB25}
 \end{figure}

We can go a step further  and  examine the circulations in each phase for $x\ge 0$ and the total  enstrophy  
 \begin{equation}
 \Gamma^{(r)}_{[0,L/2]}(t)    \equiv   \int_{x \in [0,L/2]}   \omega^{(r)}(x,y,t) dxdy,~~r=1,2;~~~~E(t) \equiv \int \int \omega^2(x,y,t) dx dy
\end{equation}
during the initial phase evolution.   It is shown in appendix \ref{RTComput} that   
\begin{equation}
  \Gamma^{(1)}_{x \ge 0}(t)    = \frac{C \sqrt{r_{\rho}}}{(1+\sqrt{r_{\rho}})}t,~~~~~\Gamma^{(2)}_{x \ge 0}(t)    =  \frac{C}{(1+\sqrt{r_{\rho}})} t,~~~~~E(t) = D~\sqrt{Re}~t^{3/2}   
   \label{MeqnIntegomegaJijVtotbisLinearizedBISfiniteampliTRISIS}
 \end{equation}  
with
 \begin{equation}
D(r_{\rho}) \equiv  \frac{ 16}{\sqrt{2}}  \frac{\sqrt{2}-1}{3\sqrt{\pi}} \sqrt{r_{\rho} \over  1 + r_{\rho}} \frac{1}{(1 + \sqrt{r_{\rho}})} \int (\Sigma_A)^2 ds
\label{initBumpenstro2TRISS}
\end{equation}
 The  scalings for circulation and enstrophy  are  respectively confirmed on figure \ref{figCircB25} and figure \ref{figEnsB25}.  The dissipation which is  equal to  $Re^{-1} \hat E $,  thus scales as  $  Re^{-1/2}$, which is indeed observed in numerical simulations.  These scalings can be useful to test  the discretization which is needed for a given Reynolds number as seen in the left 
 picture in figure \ref{figEnsB25}.

 \vspace{0.2cm}

 \subsection{Qualitative explanation for time evolution}

\begin{figure}
    \begin{center}
\includegraphics[width=0.23 \textwidth]{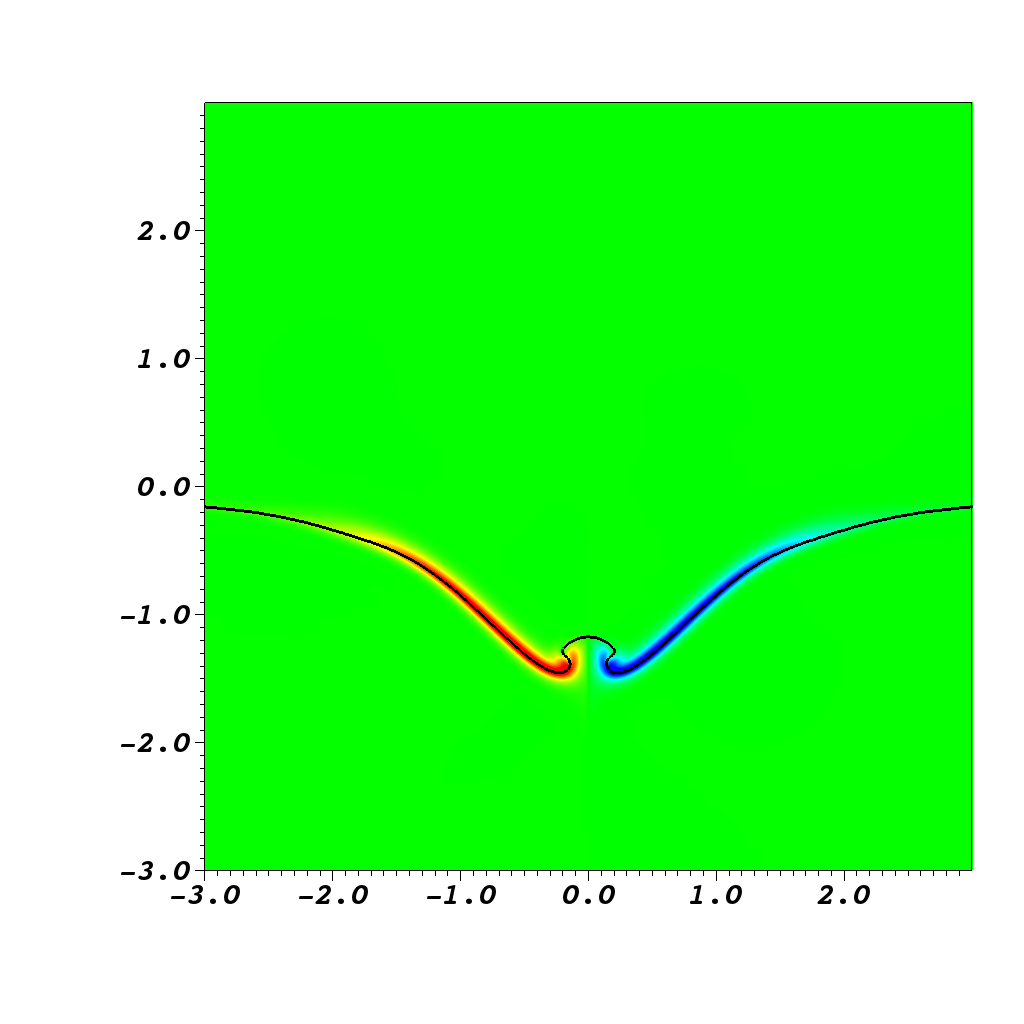}
\includegraphics[width=0.23 \textwidth]{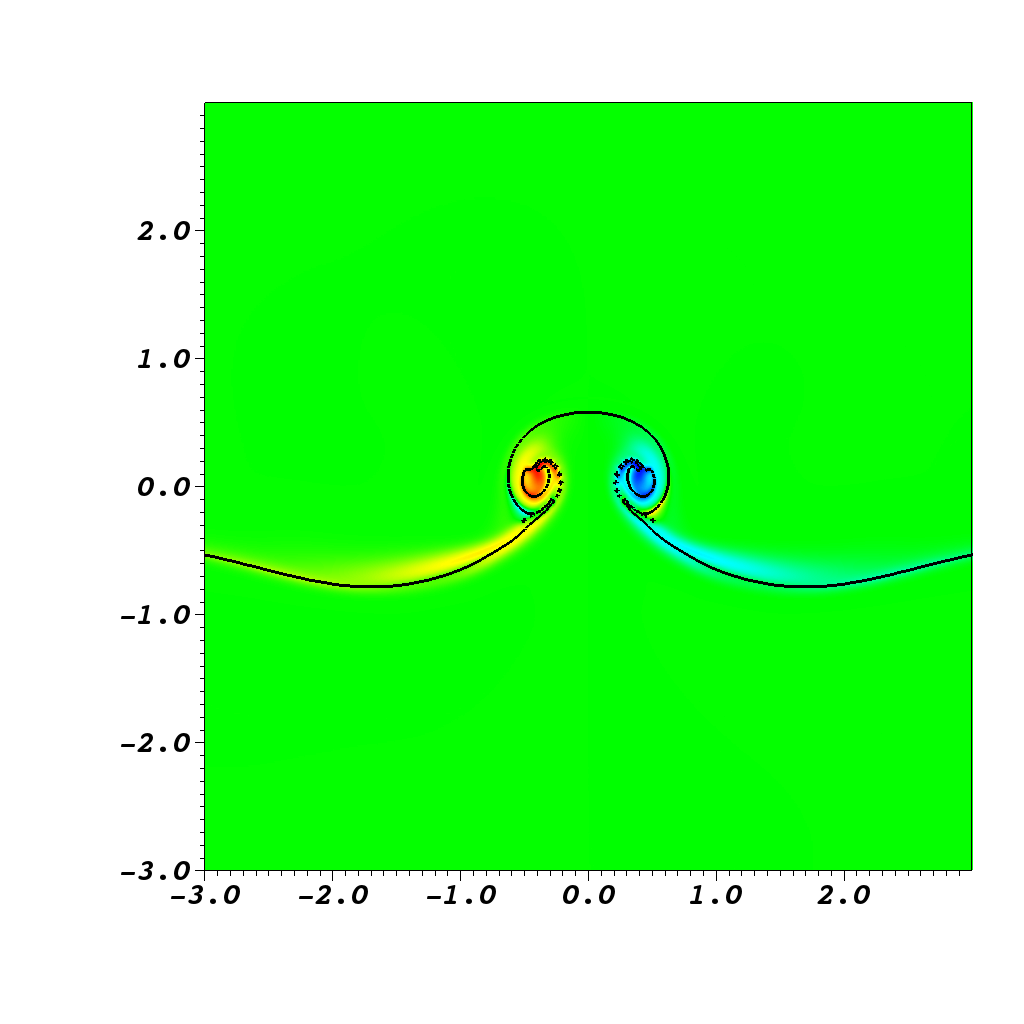}
\includegraphics[width=0.23 \textwidth]{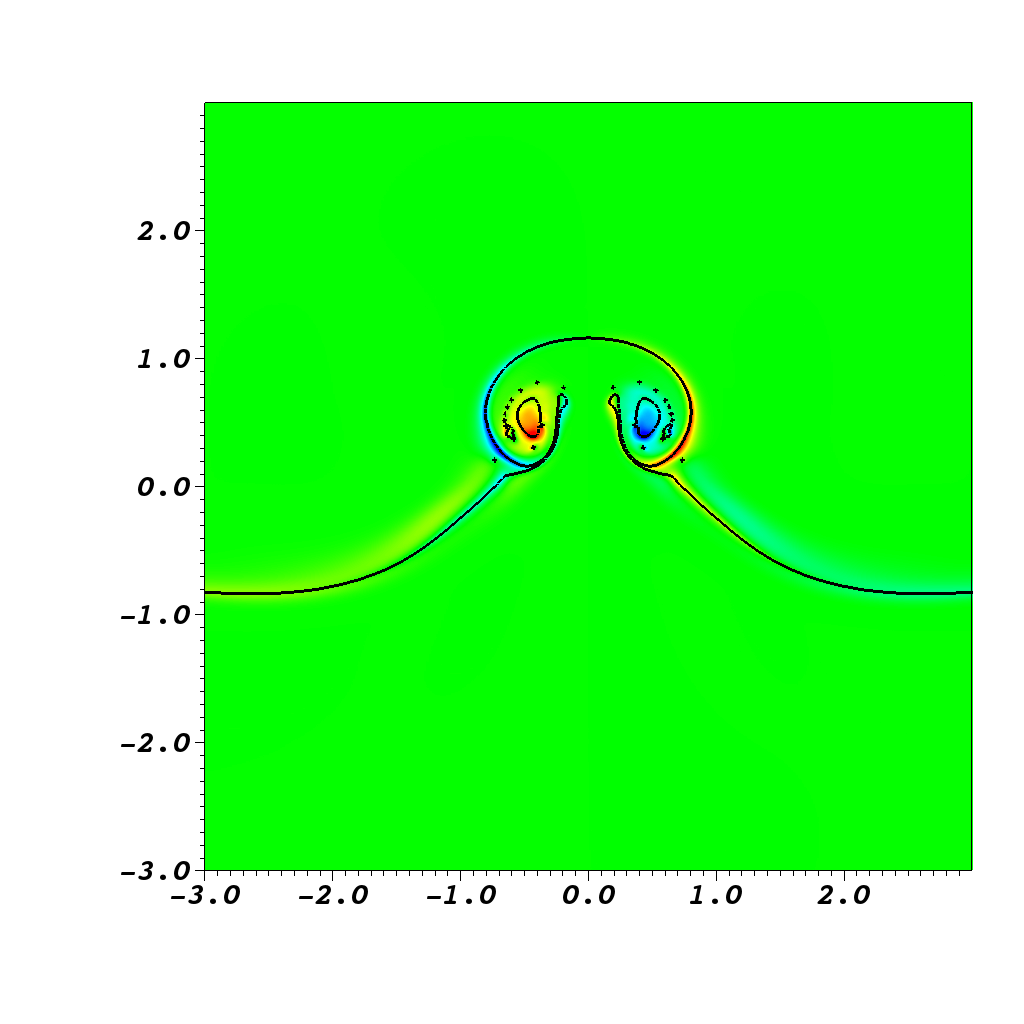}\\
\includegraphics[width=0.23 \textwidth]{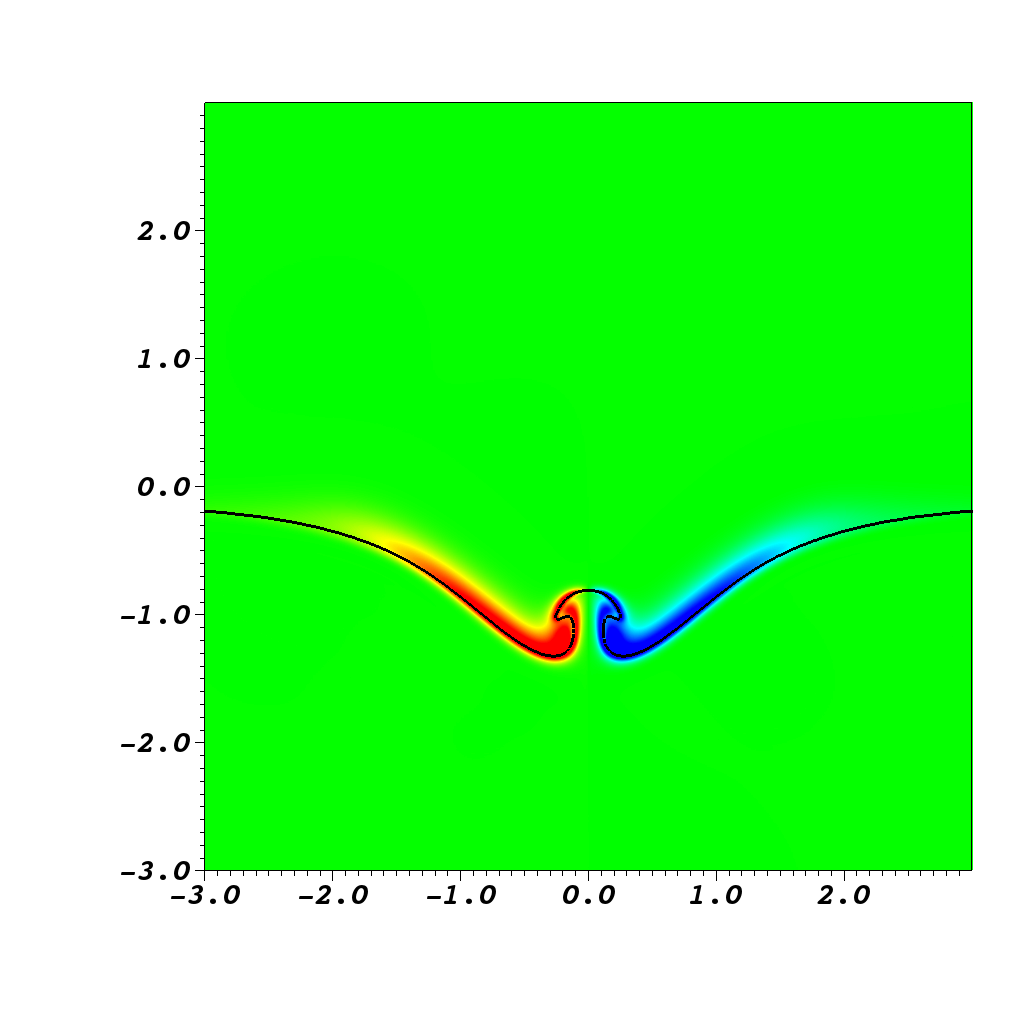}
\includegraphics[width=0.23 \textwidth]{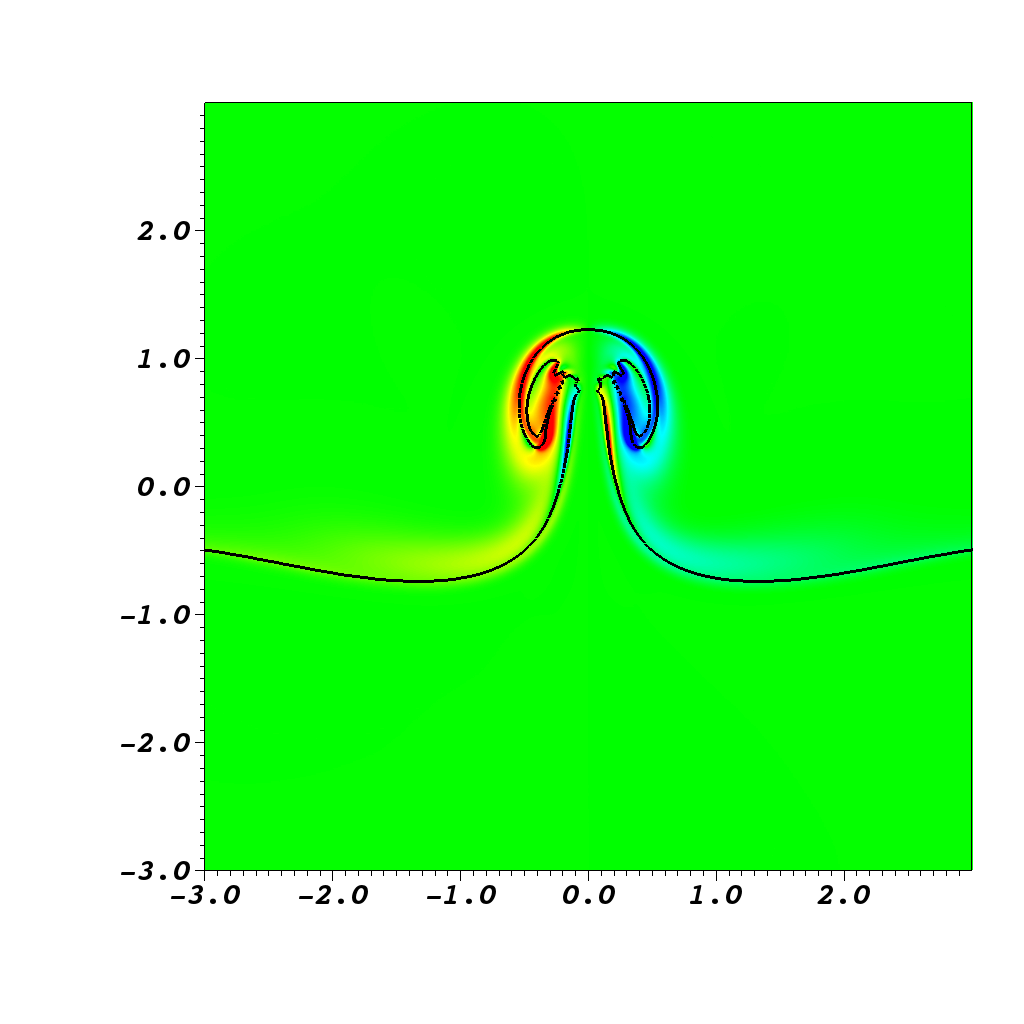}
\includegraphics[width=0.23 \textwidth]{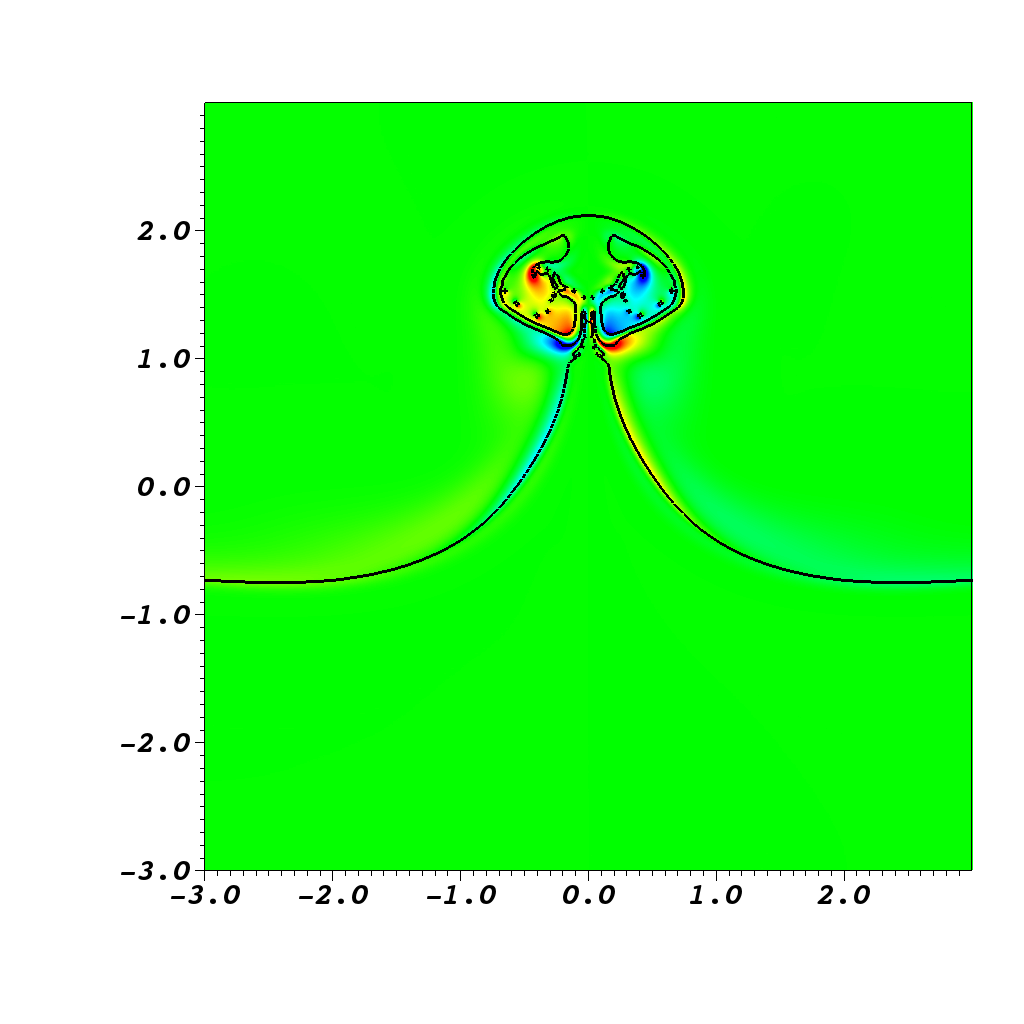}
\caption{Nonlinear   gravity perturbation characterized by  $B_0 = -2.5$, $Re=10^3$ with  (top)  $A_{tw} =1/3$ and (bottom)   
$A_{tw} =9/11$ :
Snapshots of interface and vorticity field at dimensionless  times $t = 2,4,6$.}
\label{FigBumpRe1000WeinfBneg}
\end{center}
\end{figure}

\begin{figure}
    \begin{center}
\includegraphics[width=0.23 \textwidth]{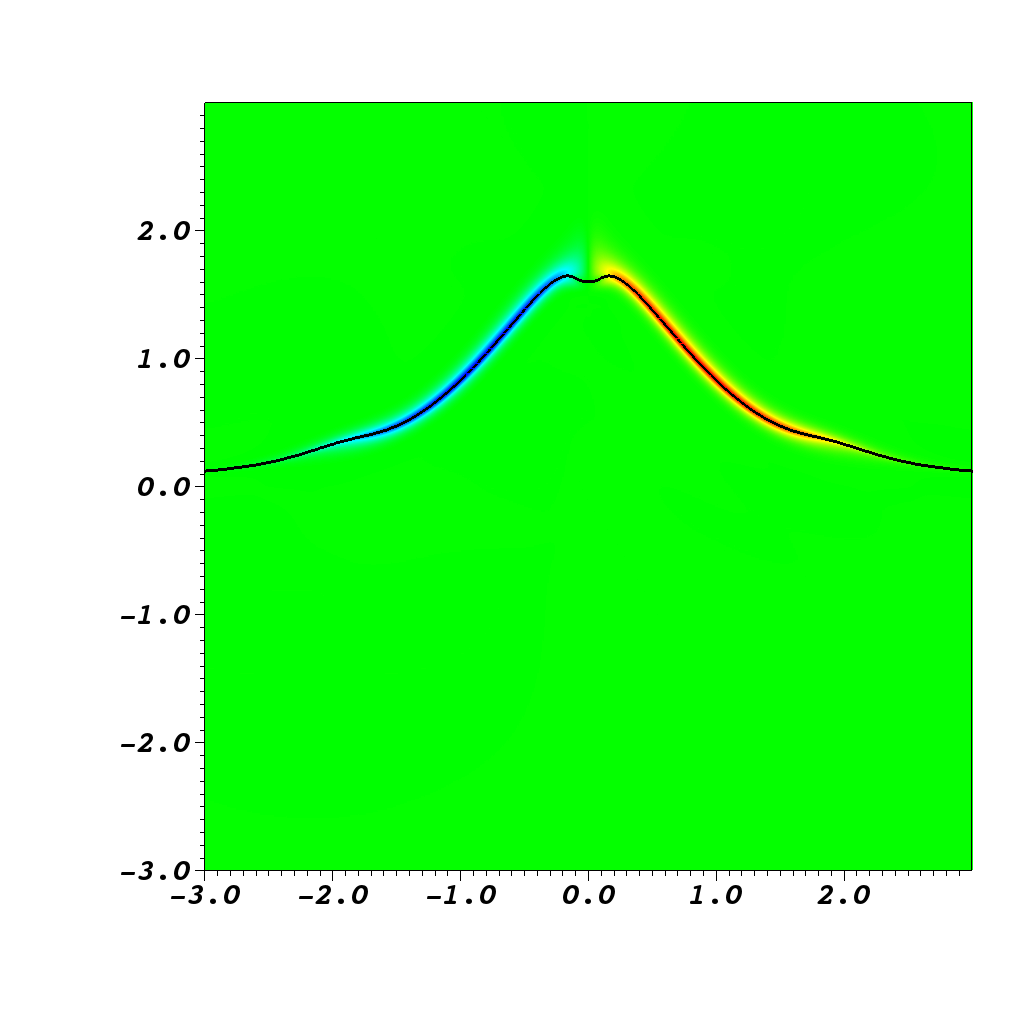}
\includegraphics[width=0.23 \textwidth]{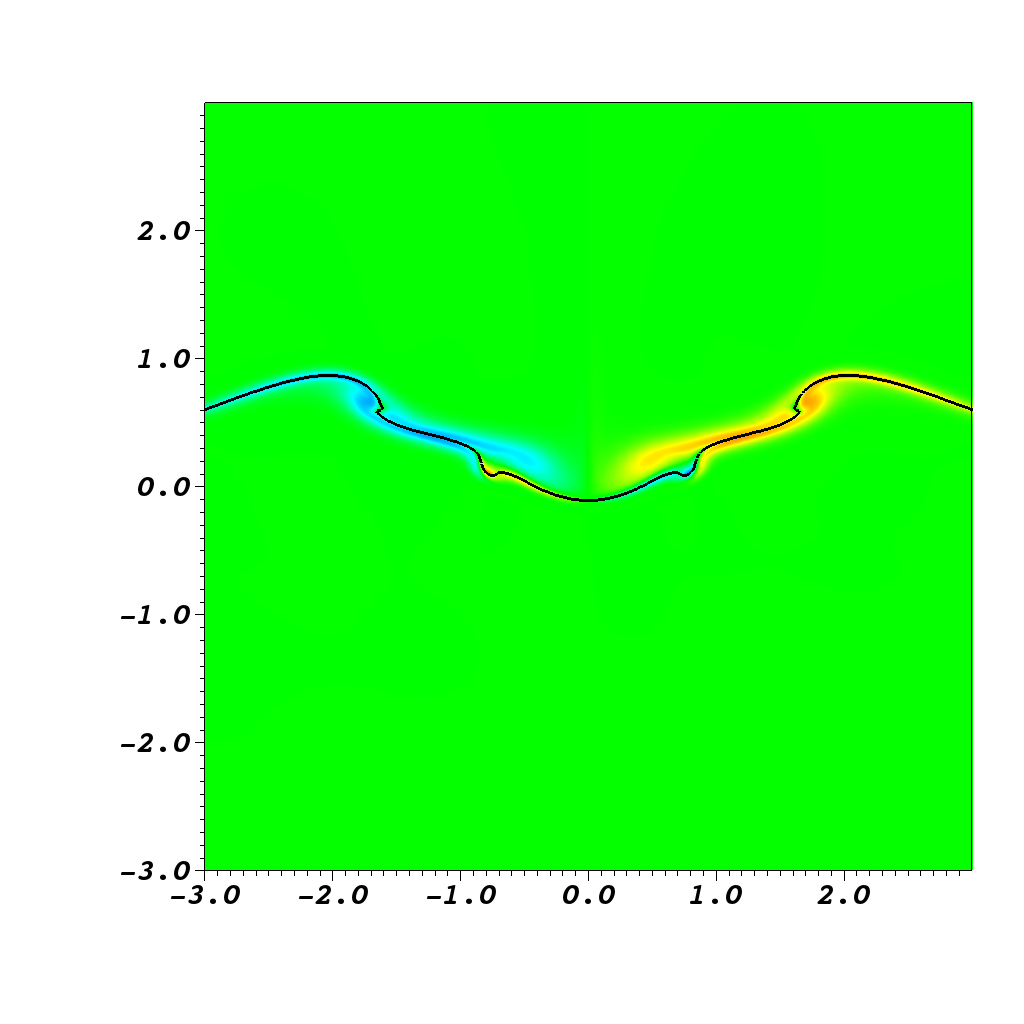}
\includegraphics[width=0.23 \textwidth]{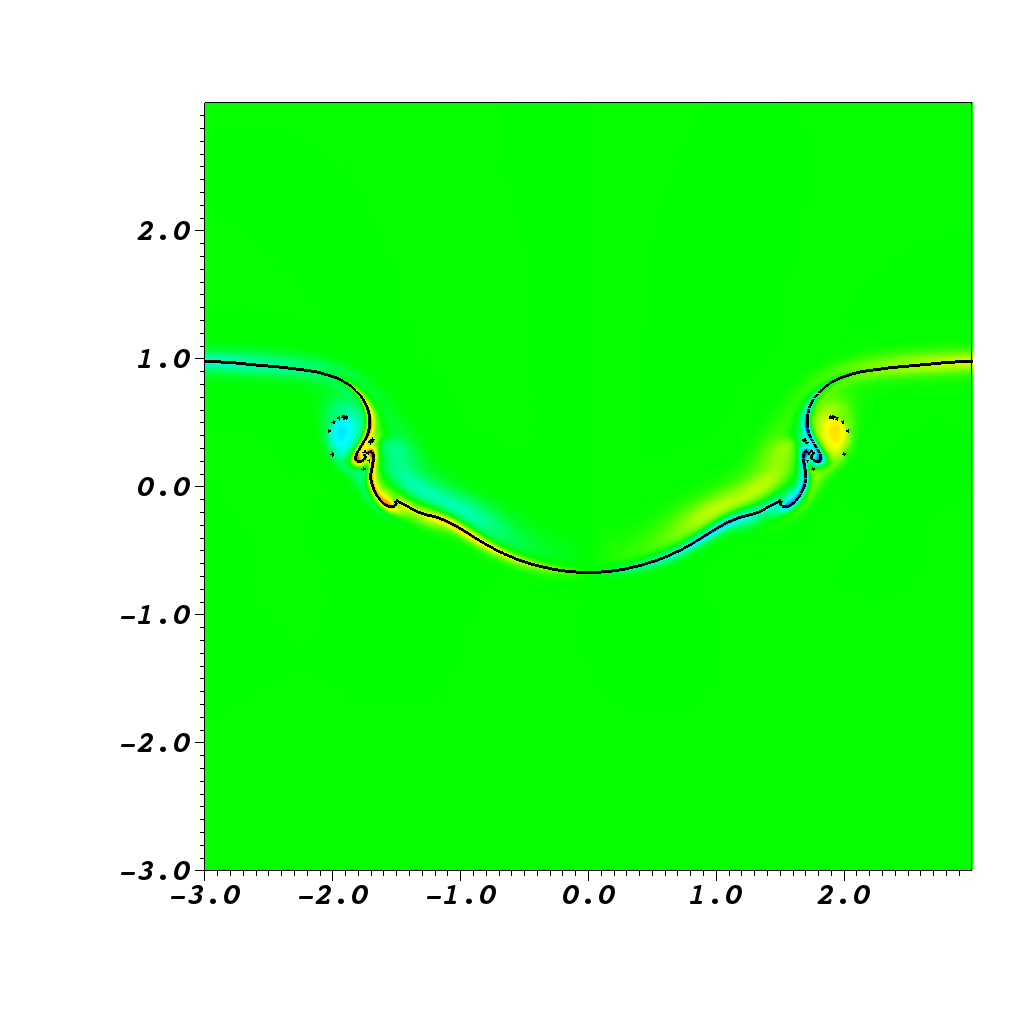}\\
\includegraphics[width=0.23 \textwidth]{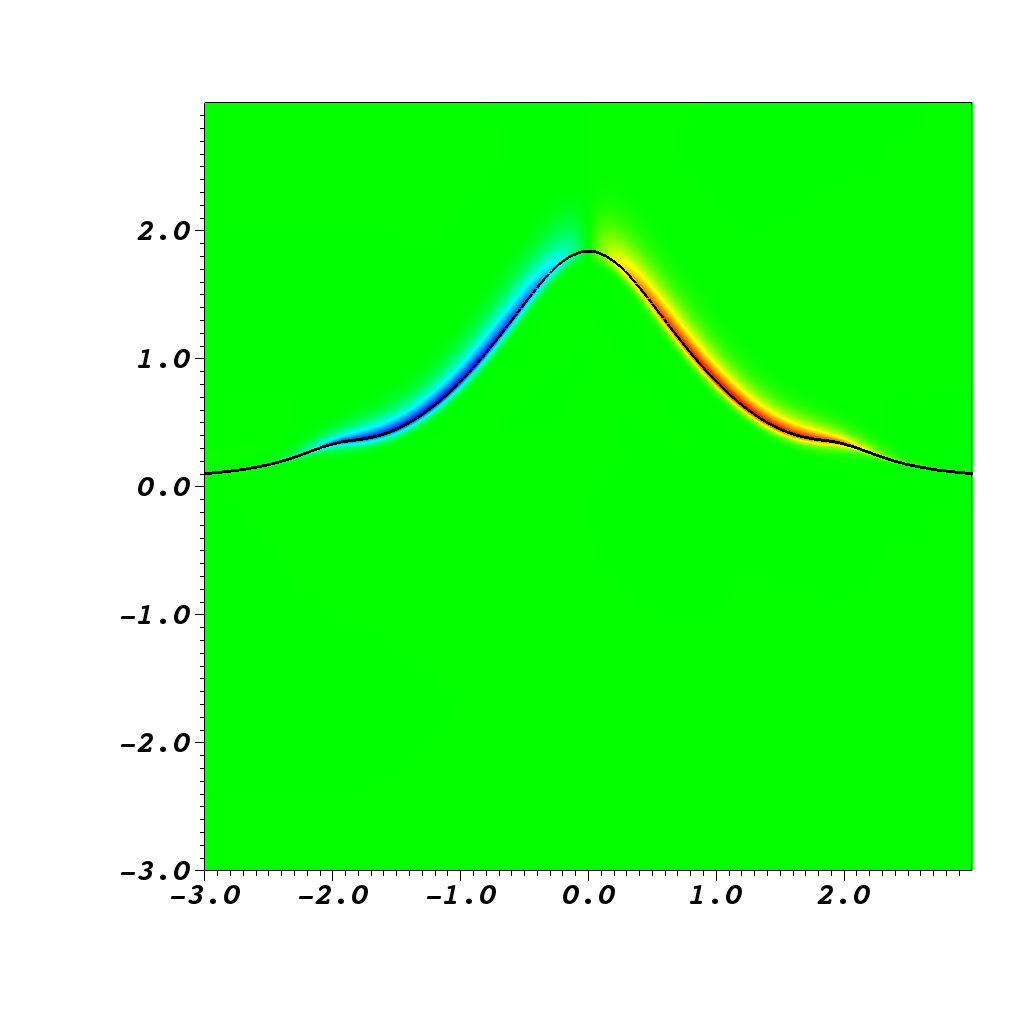}
\includegraphics[width=0.23 \textwidth]{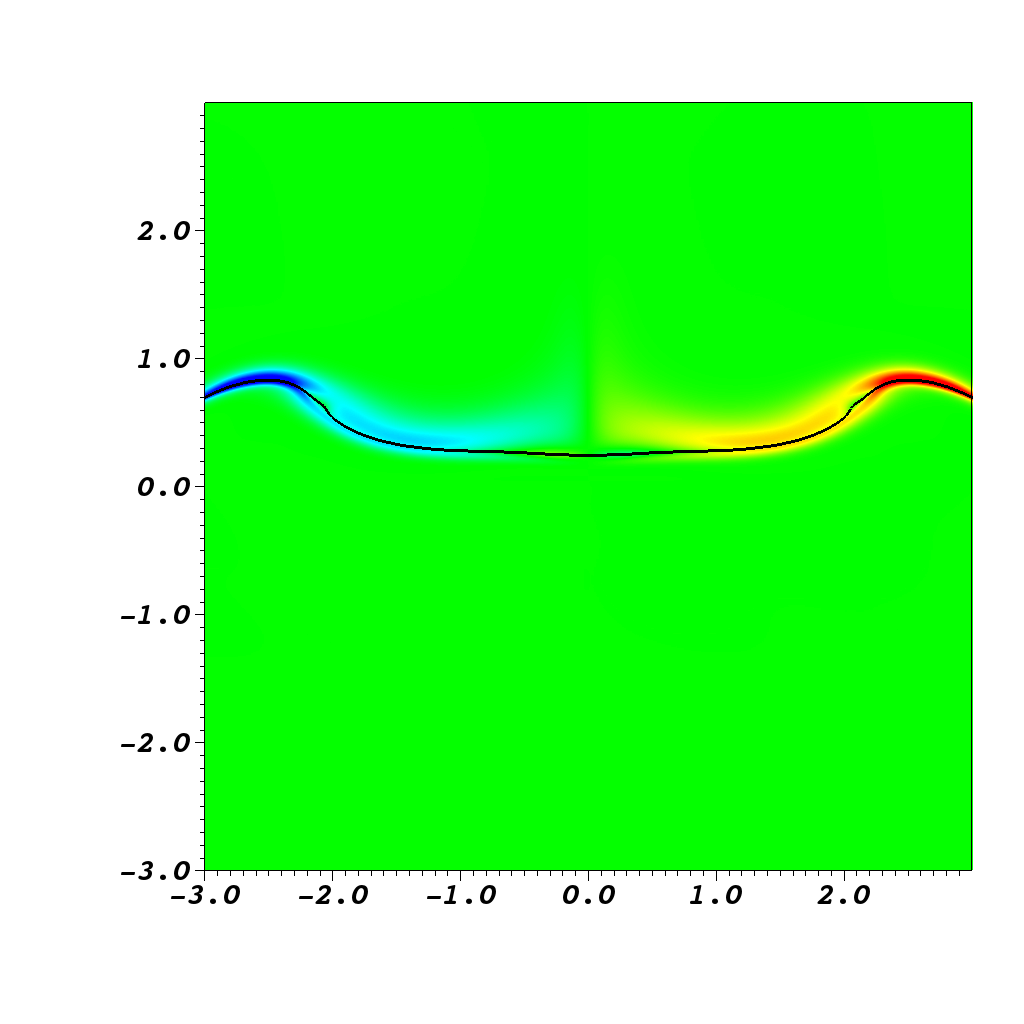}
\includegraphics[width=0.23 \textwidth]{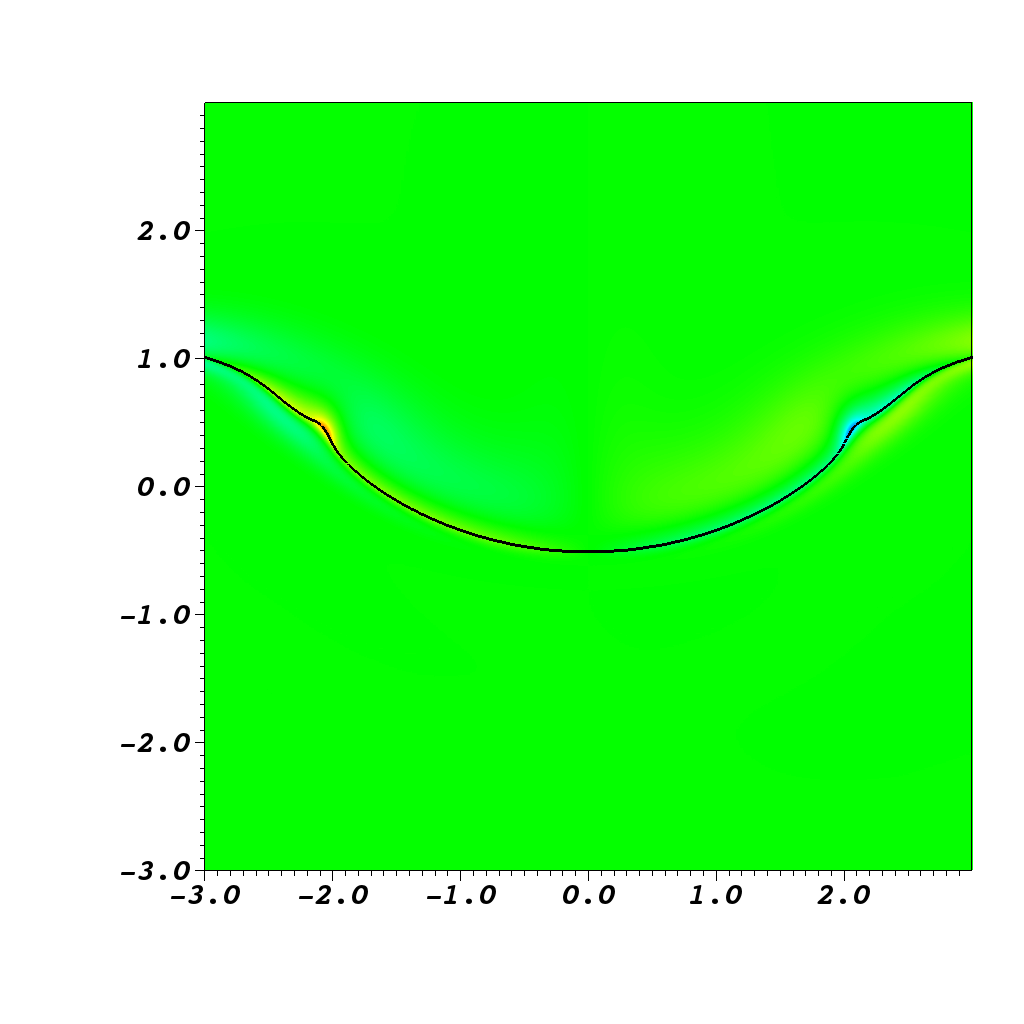}
\caption{ Idem than figure \ref{FigBumpRe1000WeinfBneg} but for    $B_0=2.5$.}
\label{FigBumpRe1000WeinfBpos}
\end{center}
\end{figure}

\begin{figure}
\begin{center}
\includegraphics[width=0.23 \textwidth]{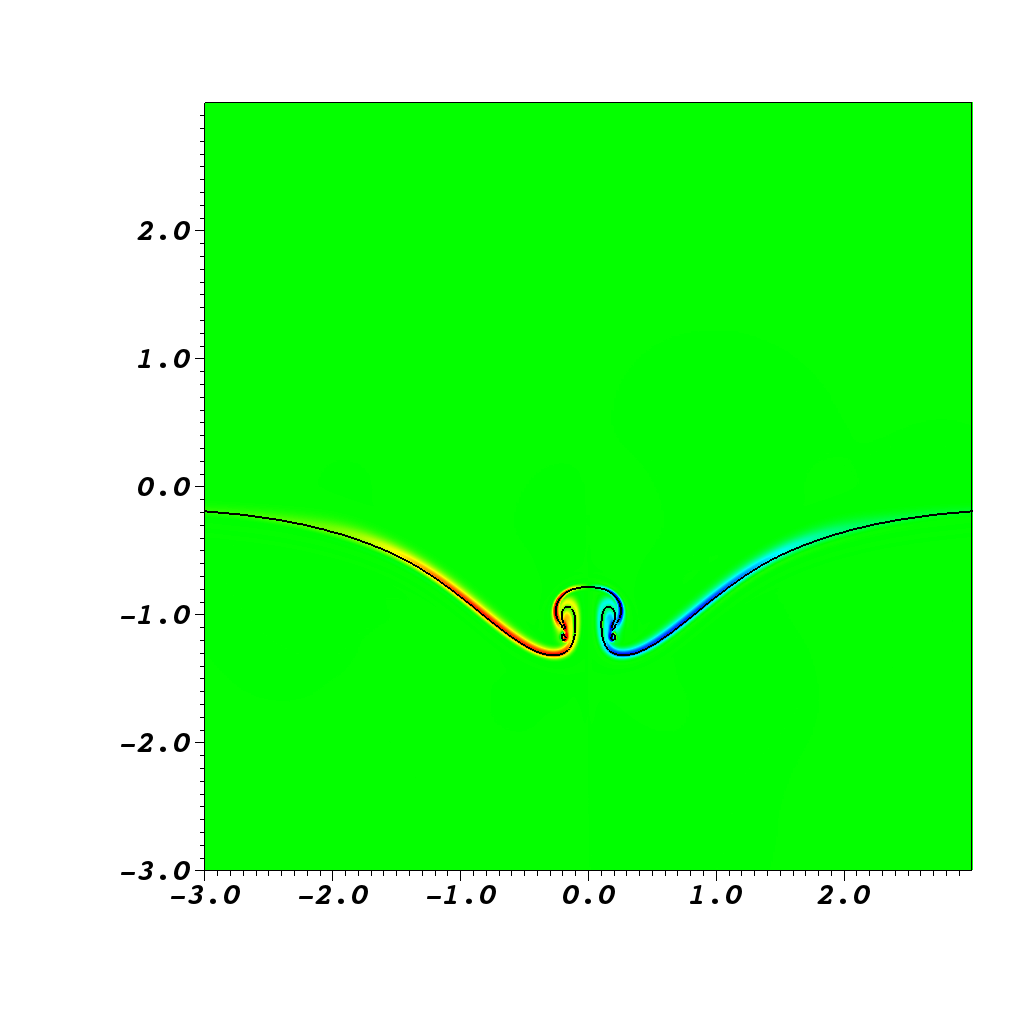}
\includegraphics[width=0.23 \textwidth]{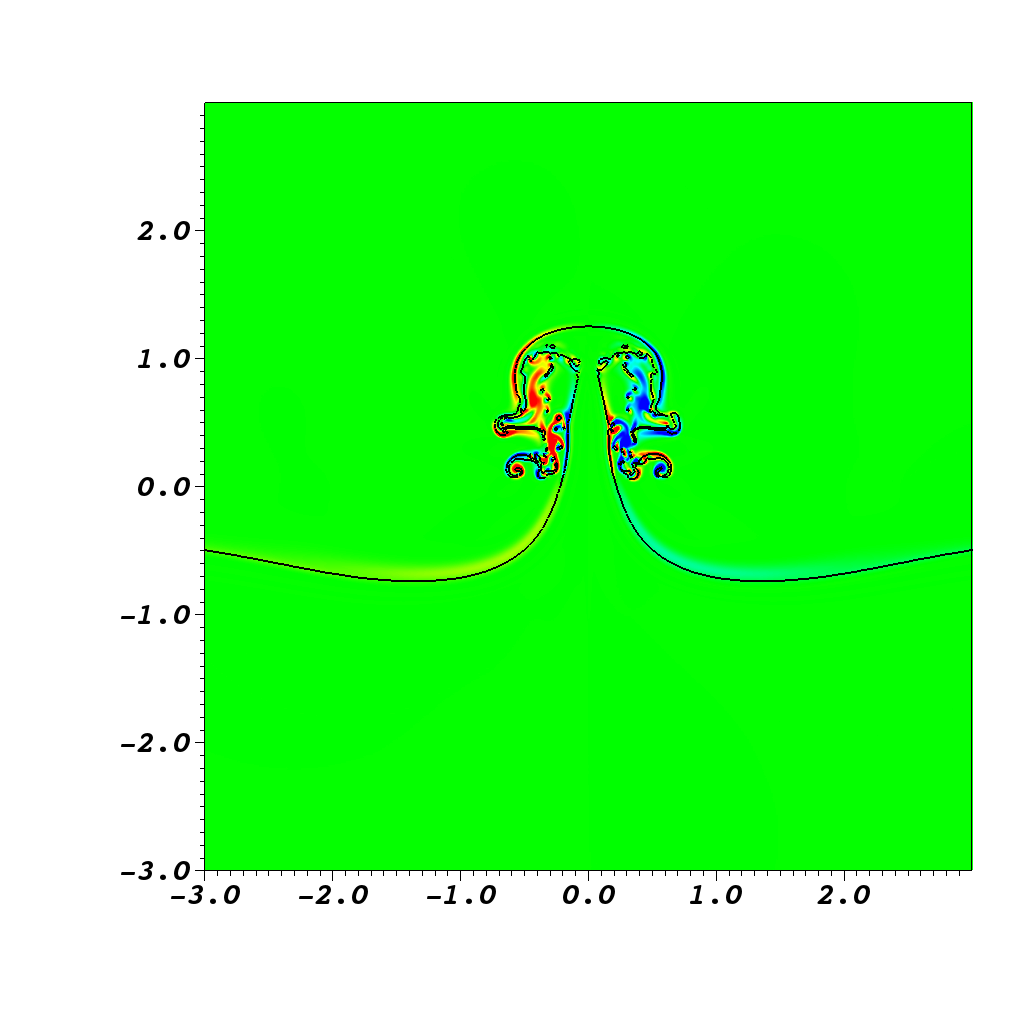}
\includegraphics[width=0.23 \textwidth]{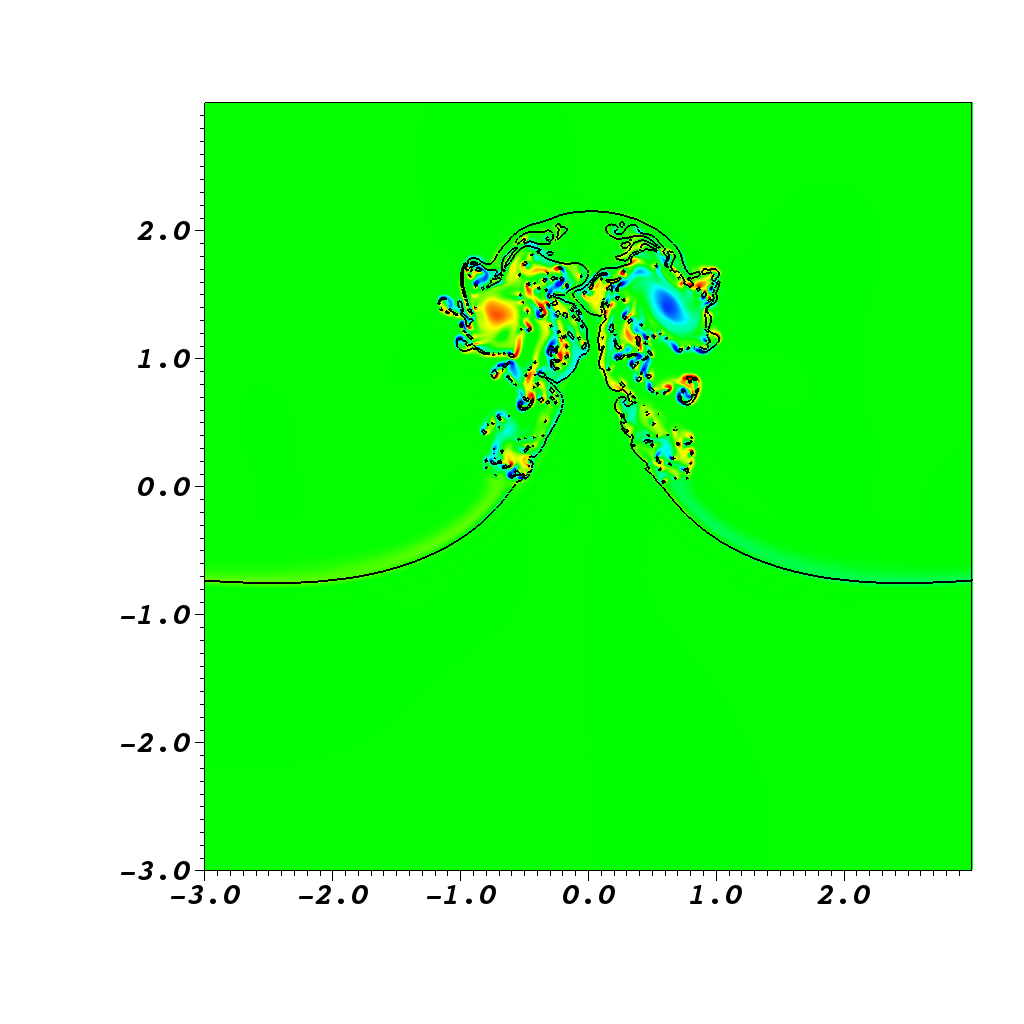}
\caption{ Nonlinear   gravity perturbation characterized by  $B_0 = -2.5$, $Re=10^4$ and $A_{tw} = 9/11$ : Snapshots of interface and vorticity field at dimensionless  times $t  = 2,4,6$.}
\label{FigBumpRe10000WeinfBneg}
\end{center}
\end{figure}

\begin{figure}
\begin{center}
\includegraphics[width=0.23 \textwidth]{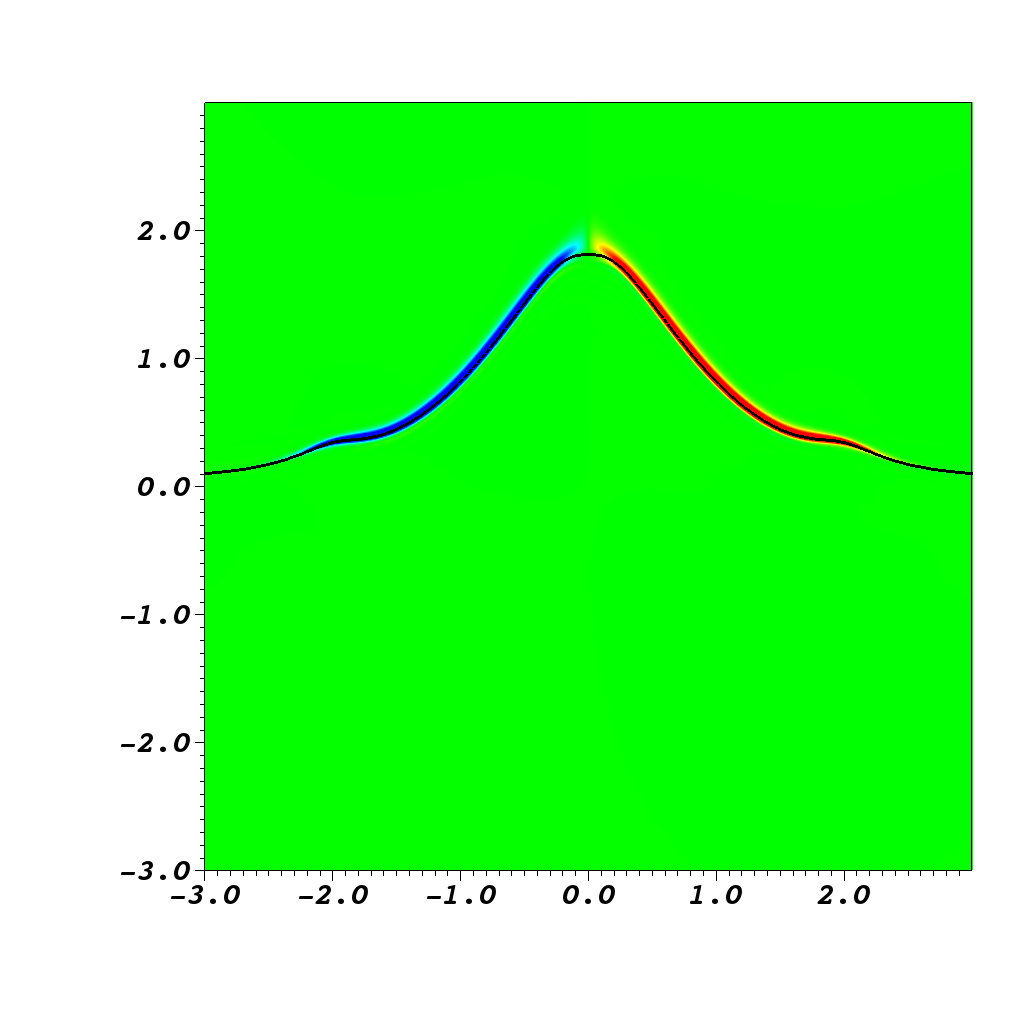}
\includegraphics[width=0.23 \textwidth]{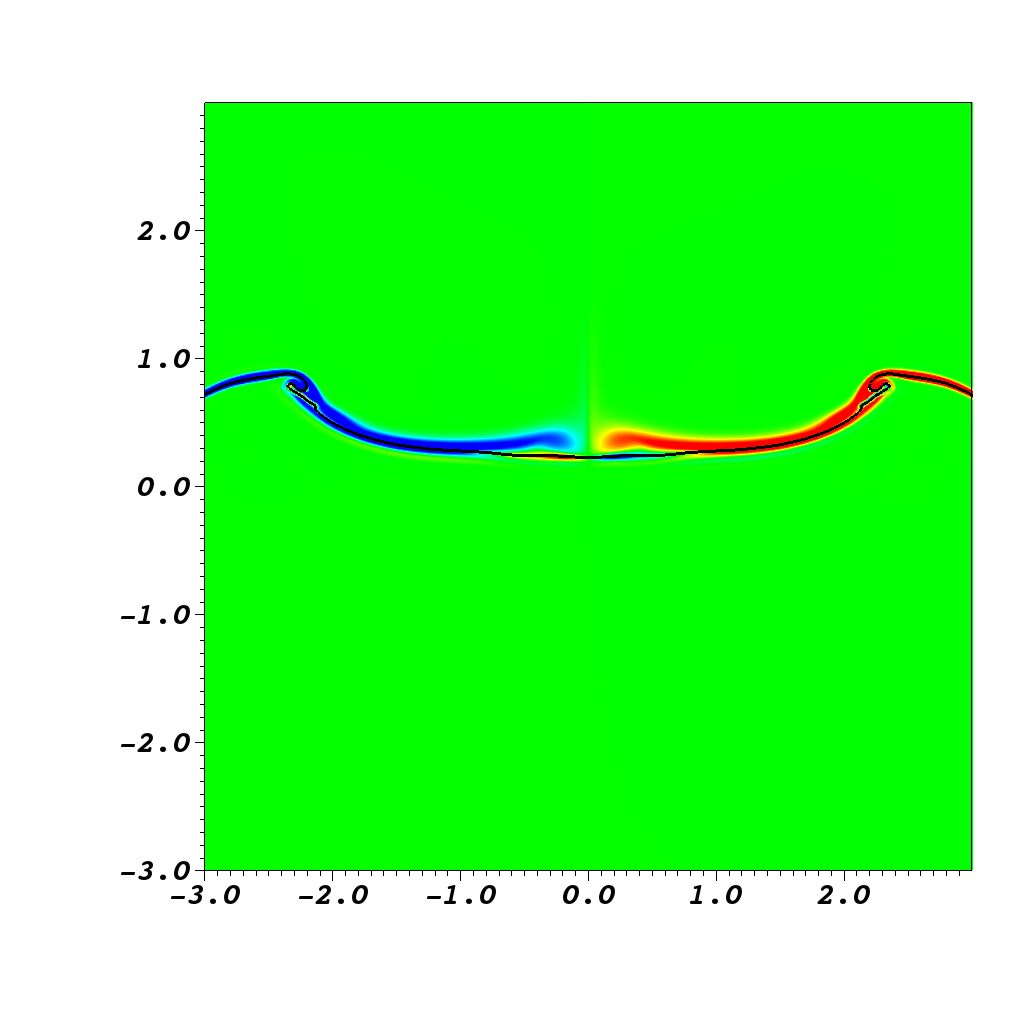}
\includegraphics[width=0.23 \textwidth]{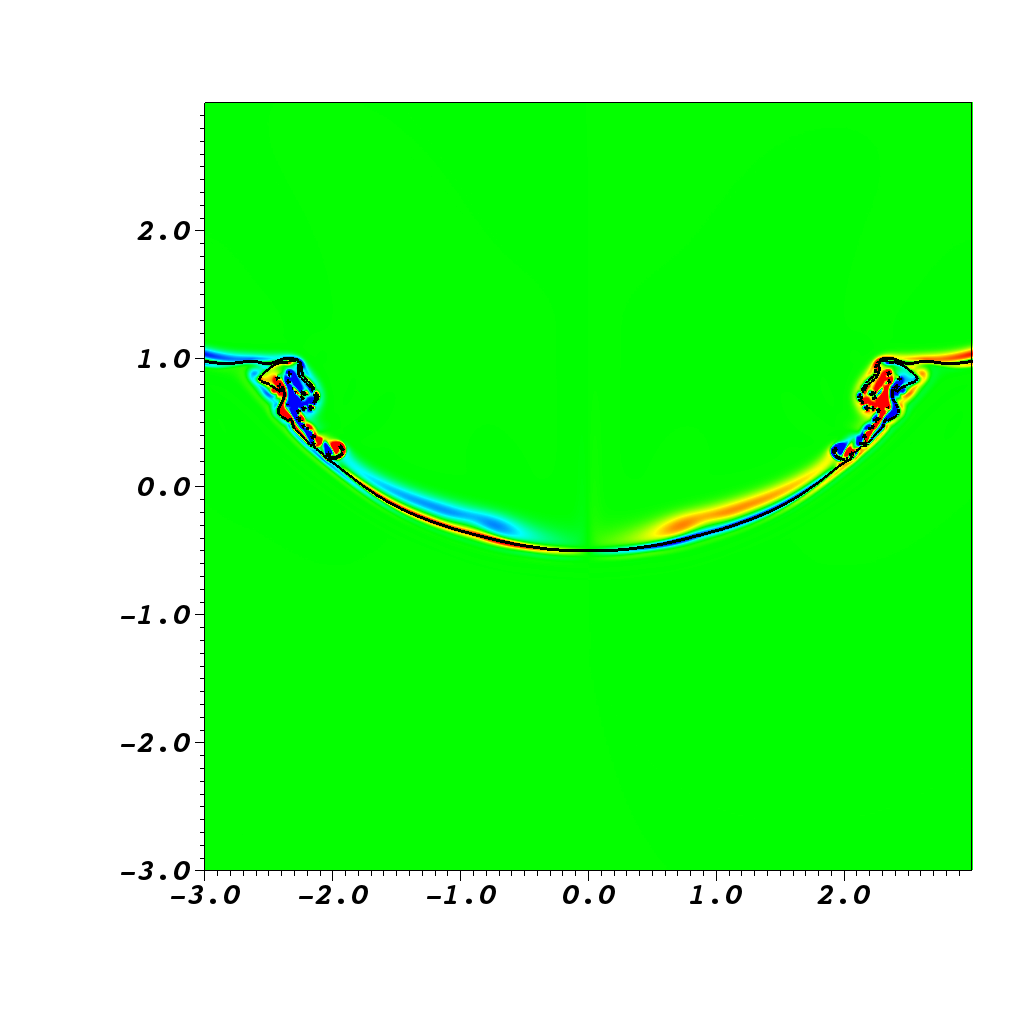}
\caption{Idem than figure \ref{FigBumpRe10000WeinfBneg}   but for    $B_0=2.5$. }
\label{FigBumpRe10000WeinfBpos}
\end{center}
\end{figure}

We   focus now    on the interface patterns. For $B_0<0$, the heavier fluid pushes the light phase  forming a  mushroom pattern (figure \ref{FigBumpRe1000WeinfBneg}). The shape depends on the density ratio  $r_{\rho}$~: the mushroom width decreases   for    increasing density ratio  (figure  \ref{FigBumpRe1000WeinfBneg}).    For $B_0>0$,    the lighter fluid tends to penetrate   the heavier one  but  the interface remains  much flatter creating a crater-like structure (figure  \ref{FigBumpRe1000WeinfBpos}).   
This problem  behaves   differently with respect to the Reynolds dependency.  
For $B_0<0$,    the Reynolds number (figure \ref{FigBumpRe10000WeinfBneg}) is affecting the mushroom structure.  For $B_0>0$,    the dependency    on the Reynolds number is not significant (figure  \ref{FigBumpRe10000WeinfBpos}).
 The interface evolution  is  induced by the vorticity generated on the
 interface itself  which in turns depends on the source.   However one may
 qualitatively  justify  the different behaviours observed  by noting which
 source terms are dominant during each phase.  When ${\rho^{(2)} } /
 \rho^{(1)}>>1$, we use  equation \eqref{MeqnSOmegaInterfaceNEWnondimensional} where we neglect the Reynolds
 part (we only consider problems where viscosity has a small influence on the
 total production rates, which is valid in the limit of infinite Reynolds) so
 that    
 \begin{equation}
  \hat \Psi_{\Sigma}  \approx {\hat \Psi}_{pot}+ {\hat \Psi}_{\rho_m} ~~~~\hbox{with}~~~ \hat {\Psi}_{pot}=- \frac{\hat \eta}{\pi}.
   \label{MeqnSOmegaInterfaceNEWbisnew}
\end{equation}
Near $t=0$  the potential term ${\Psi}_{pot}$ is significant  and the velocity
term ${\Psi}_{\rho_m}$ is  negligible. By contrast, when the interface amplitude
becomes weaker and velocity is sufficiently large,  ${\Psi}_{\rho_m}$  is dominant and ${\Psi}_{pot}$ negligible.
Let us  assume then a two step process. In a first period,    the gravity source term ${\Psi}_{pot}$     only  produces  the vorticity field.  In a second  period,   this vorticity field is modified by the total source.
Using the approximate expression \eqref{MeqinitBumpinterface1append}    and assuming it to be valid  similar to the principal Fourier mode $k=\pi$ (this mode amplitude  goes to zero at time    $T_f=\pi/(2\hat \varpi) $ for a pulsation $\hat \varpi= 1$) then 
at $t=\frac{\pi}{2}$:
\begin{equation}
 \omega(x,y,t) =   
    \begin{cases}
     \omega^{(1)}(x,y)  =  - 2\pi\frac{\Sigma^{(1)}_A(x)}{\sqrt{2}\delta^{(1)}}  G (\frac{y}{\sqrt{2}\delta^{(1)}})~~& \text{if  $ y \ge 0$}\\
       \omega^{(2)}(x,y) =   -2\pi \frac{\Sigma^{(2)}_A(x) }{\sqrt{2}\delta^{(2)}}  G (-\frac{y}{\sqrt{2}\delta^{(2)}})~~ & \text{if $y \le 0  $}
    \end{cases}       
\label{MeqFlatinterface1}
\end{equation}
with
$$
 G(z) \equiv  -\frac{1}{\sqrt{\pi}}   \exp (-z^2)    + z\bigg[1- Erf(z) \bigg]
$$
and  
\begin{equation}
 \delta^{(1)}=    \sqrt{1+r_{\rho}} \sqrt{\frac{\pi}{2Re}},~~~~~
 \delta^{(2)}=      \sqrt{\frac{1+r_{\rho}}{ r_{\rho}} }\sqrt{\frac{\pi}{2Re}}
\label{MeqinitBumpinterface2append}
\end{equation}
\begin{equation}
\Sigma^{(1)}_A(x)=\frac{\sqrt{r_{\rho}}}{1+\sqrt{r_{\rho}}}   \Sigma_A(x), ~~
\Sigma^{(2)}_A(x)=\frac{1}{1+\sqrt{r_{\rho}}}  \Sigma_A(x)
\label{initBumpinterface3append}
\end{equation}
\begin{equation}
\Sigma_A(x) =   \frac{4B_0}{\pi}  x  \exp(-x^2) 
\end{equation}
One then  initializes a new  simulation   with  two-dimensional  vorticity field generated by this source.    
\begin{figure}
    \begin{center} 
\includegraphics[width=0.23 \textwidth]{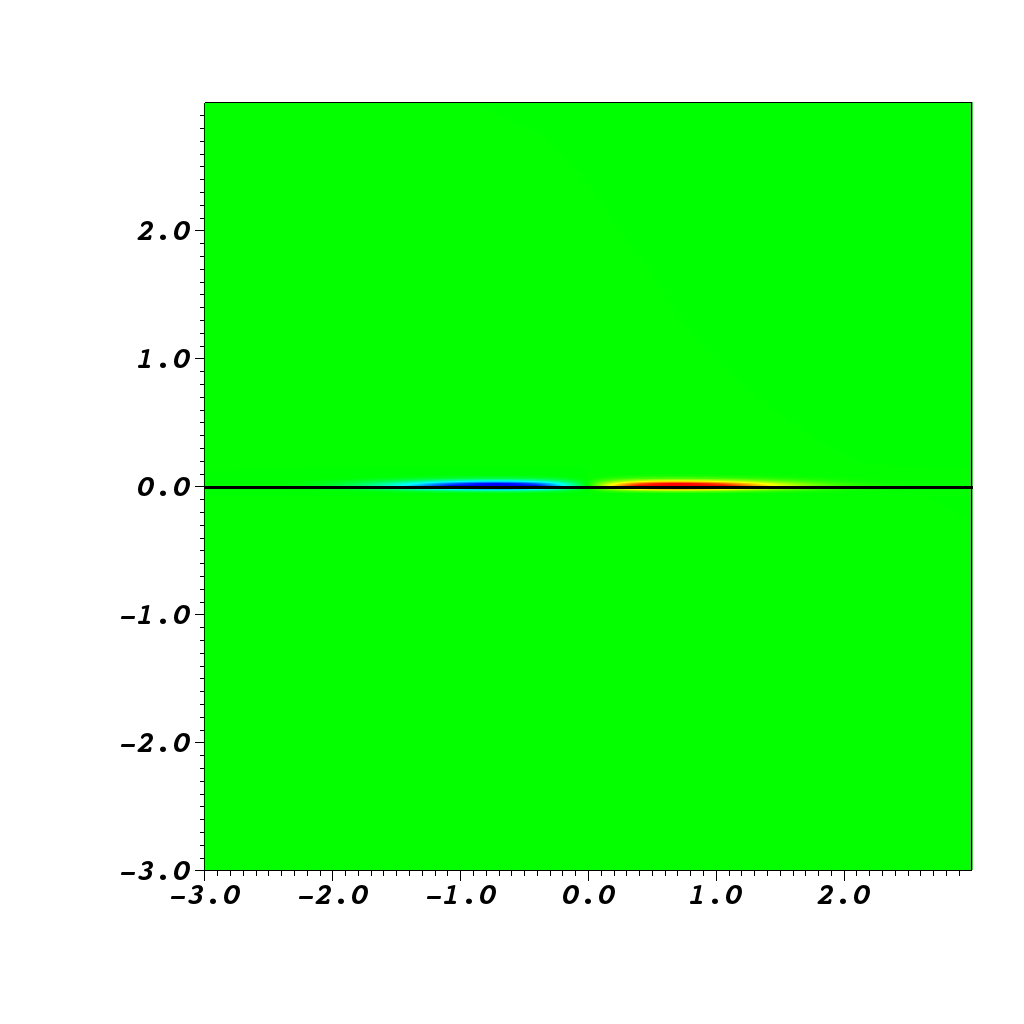}
\includegraphics[width=0.23 \textwidth]{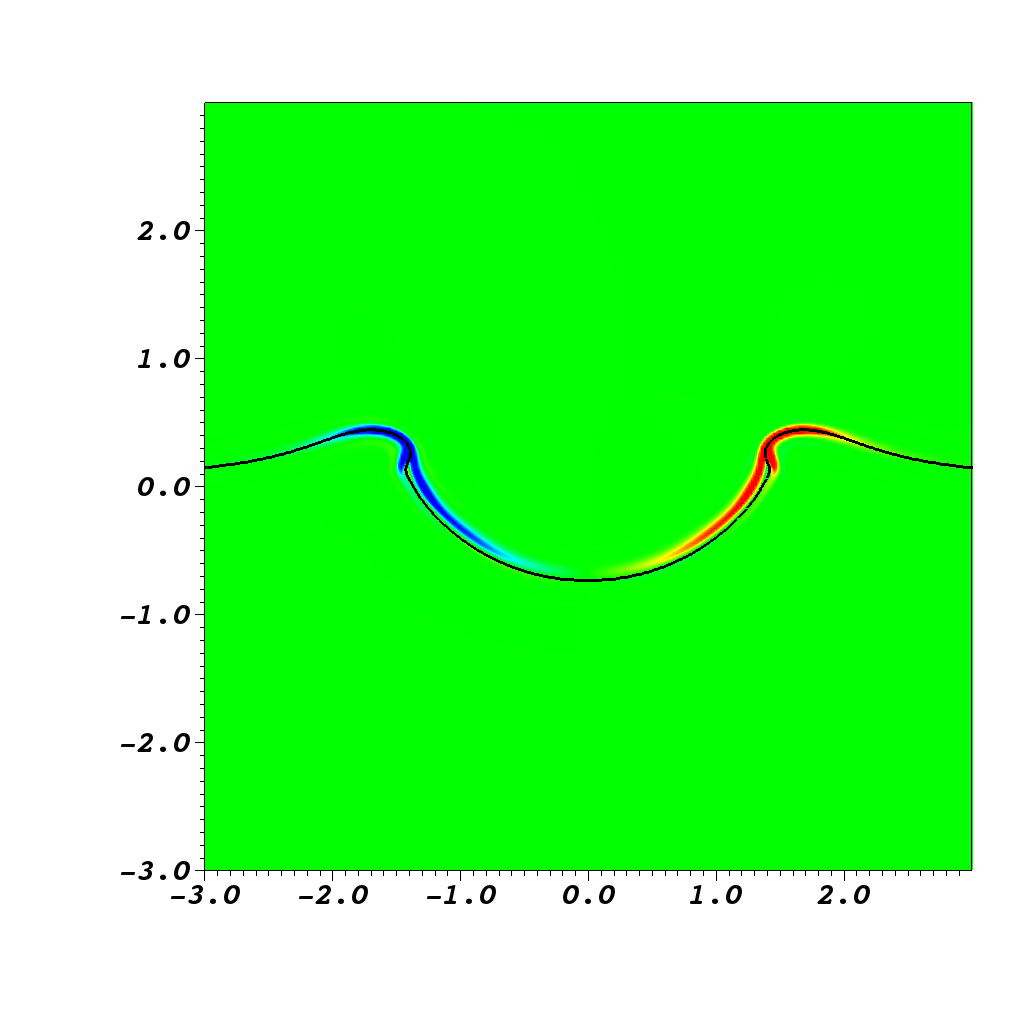}
\includegraphics[width=0.23 \textwidth]{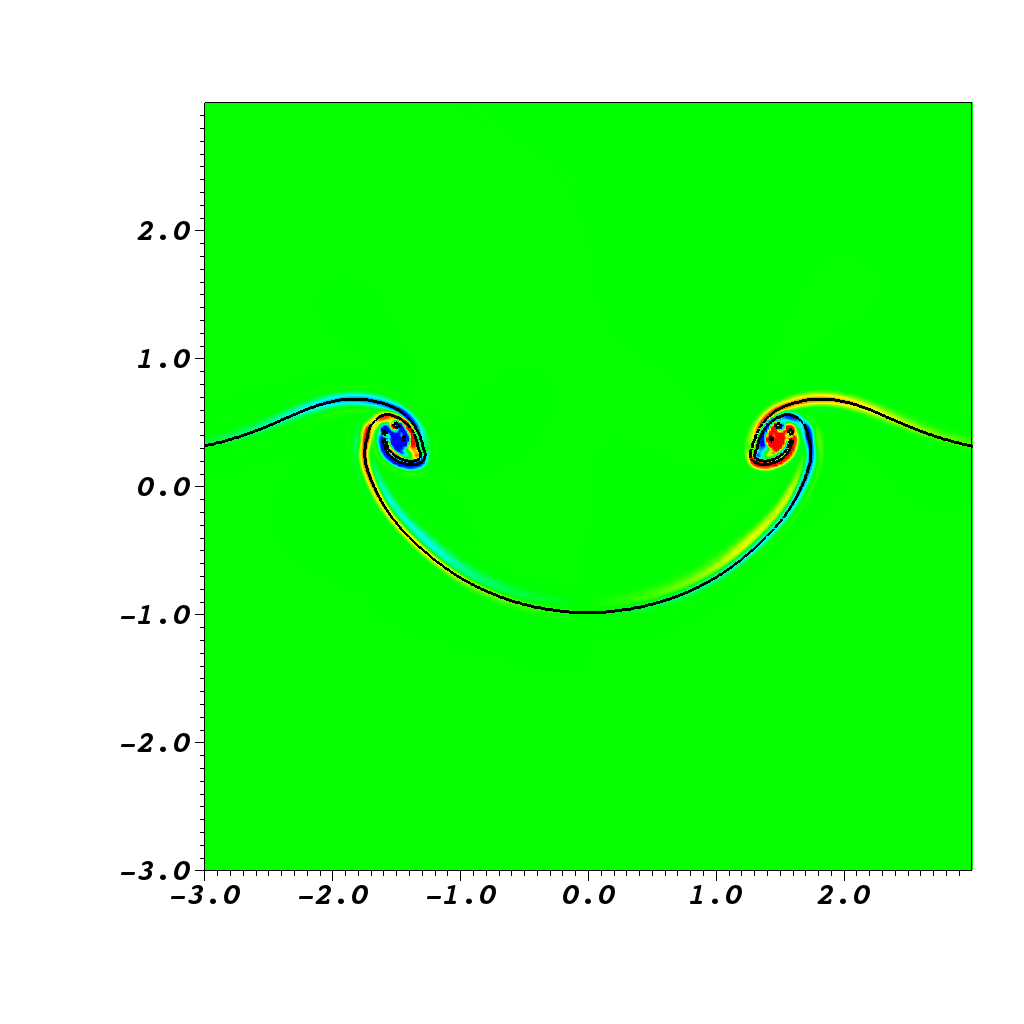}\\
\includegraphics[width=0.23 \textwidth]{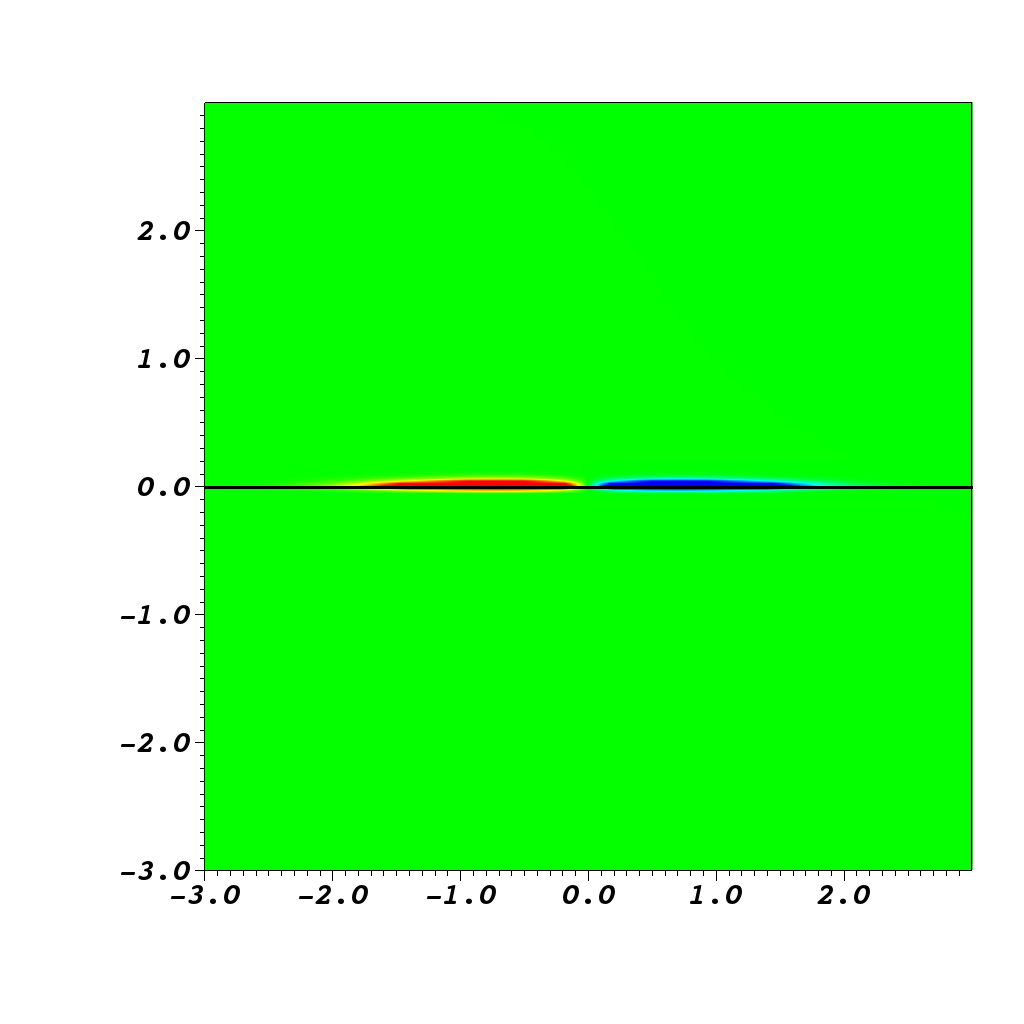}
\includegraphics[width=0.23 \textwidth]{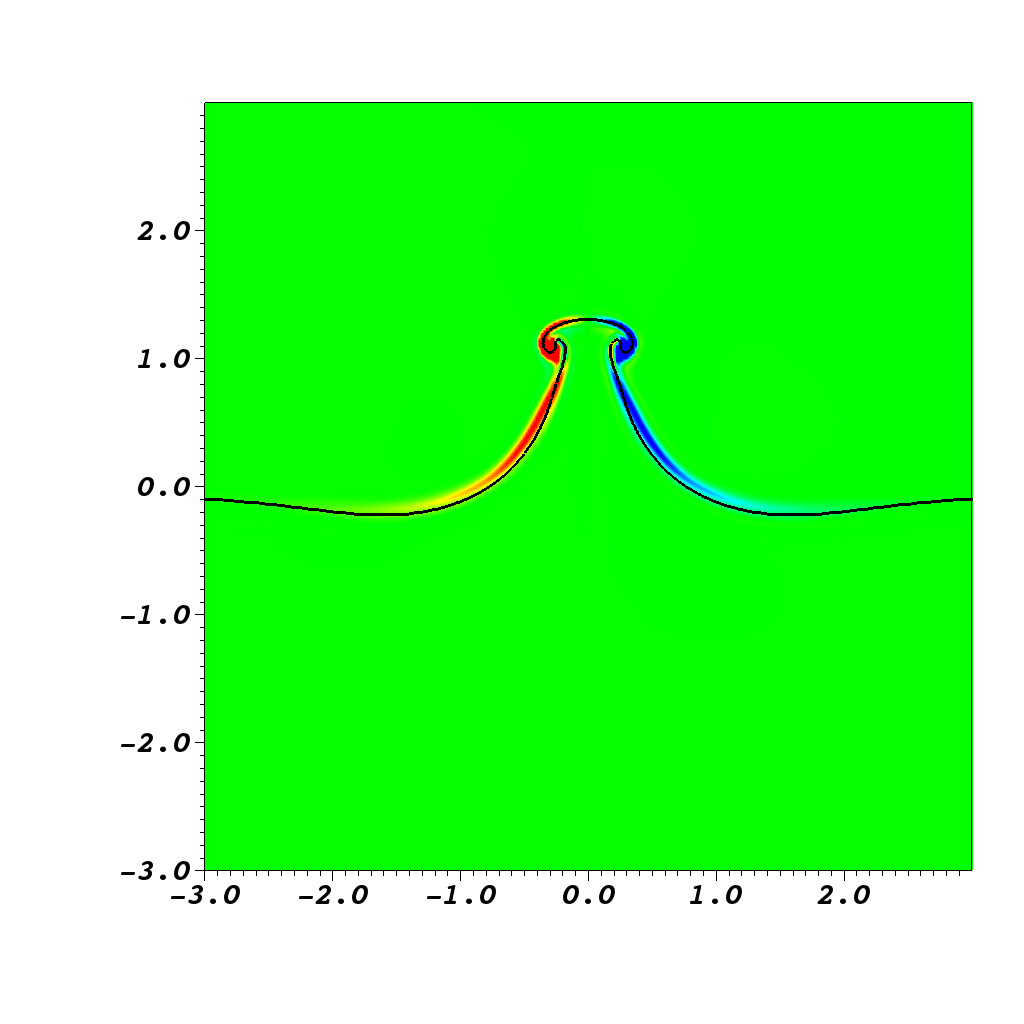}
\includegraphics[width=0.23 \textwidth]{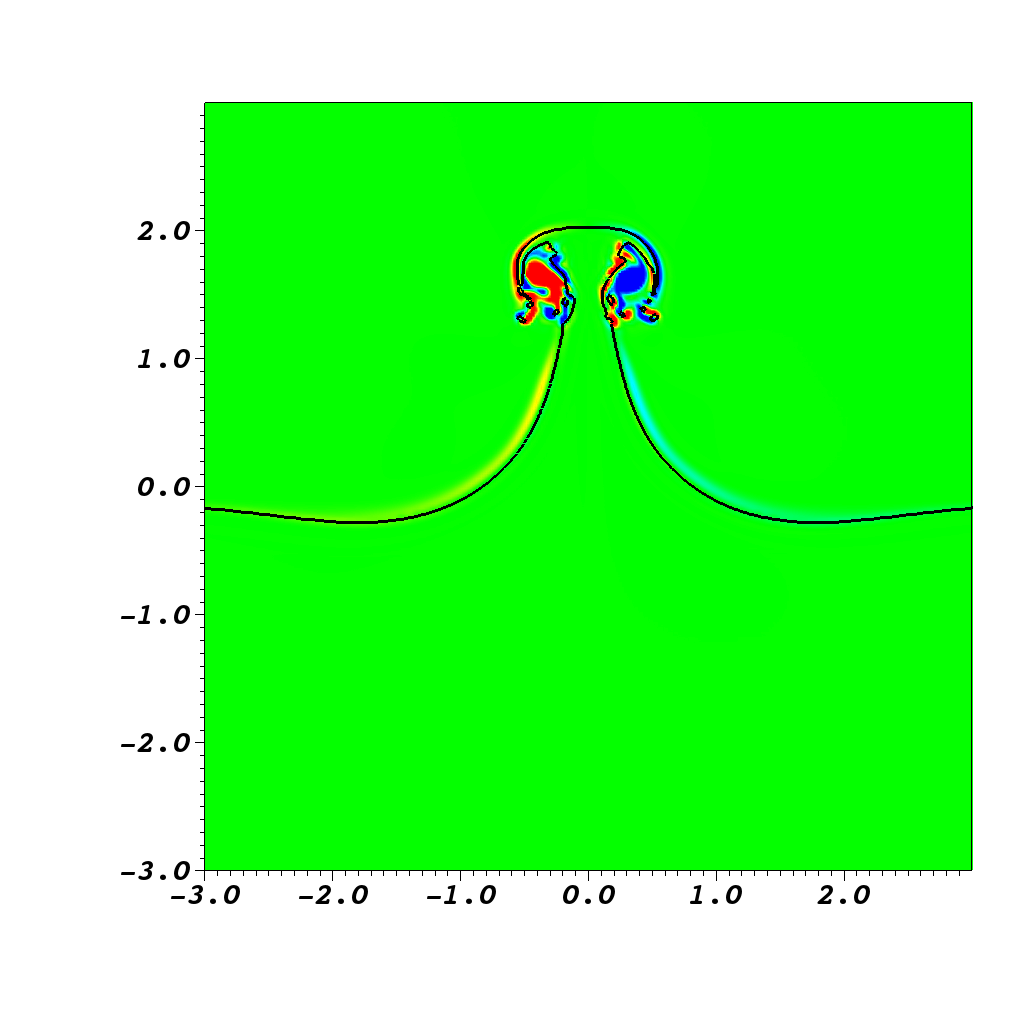}
\caption{ Nonlinear   gravity perturbation characterized by  $A_{tw} = 9/11$,  $Re=10^4$ and (top) $B_0=2.5$,  (bottom) $B_0=-2.5$: evolution of an initial flat interface at time $t=0,1,2$.
An initial vorticity  layer is present  on the interface corresponding to equation \eqref{MeqFlatinterface1}.
}
\label{figBumpR10WeinfFLATgammaBpositive}
\end{center}
\end{figure}
 The numerical simulations show that  whether we reproduce the  positive bump case   
 or the  negative bump case  
  similar structures than those observed in the original problem 
  are found(figure \ref{figBumpR10WeinfFLATgammaBpositive}): in the case of $B_0 < 0$ a jet similar to that of figure \ref{FigBumpRe1000WeinfBneg} is observed, whereas for $B_0>0$ a crater like structure similar to that of figure \ref{FigBumpRe1000WeinfBpos} appears. To explain this non-symmetric behavior let us   compute  the source   the numerical method of appendix \ref{AppendixNumericalimplementation} for a flat surface( to be called configuration B). In that instance, the  two gravitational terms (with $\eta$ and $  \Psi_{g}$) are zero and the inertial term    dominates  the vorticity flux.\\
  
\begin{figure}
\begin{center}
\includegraphics[width=0.45 \textwidth]{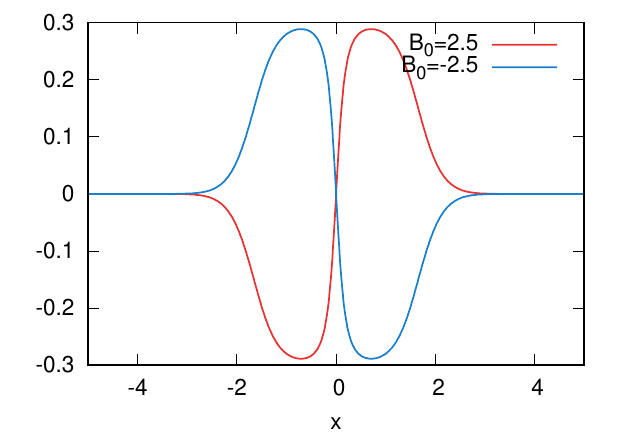}
\includegraphics[width=0.45 \textwidth]{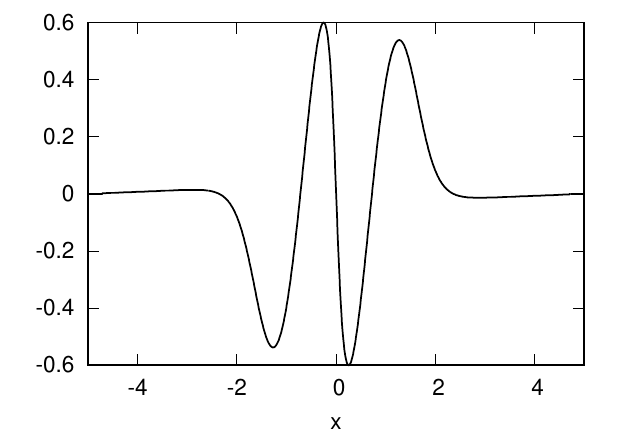}
\caption{Nonlinear   gravity perturbation characterized by  $|B_0| = 2.5$,  $Re=\infty$ and $A_{tw} \to 1$: 
Sources $\Sigma$ in (left) configuration A (bump and flow at rest) and (right) configuration  B (flat interface and velocity).
 }
\label{FigSourcesAB}
\end{center}
\end{figure}

\noindent    Figure \ref{FigSourcesAB} shows the structure of the sources in configuration A and B in the limiting case of $A_{tw} \to 1$. 
We first note that the term ${\Psi}_{pot}$ is symmetric with respect to the symmetry $\eta^*(x,t) = -\eta(x,t)$
and   $ (u^*_1(x,y,t),u^*_2(x,y,t)) =(u_1(x,-y,t),-u_2(x,-y,t))$,  while the term
${\Psi}_{\rho_m}$ is said to be antisymmetric, as the source has the same sign irrespective of the sign of the perturbation $B$.
This has important consequences on the vorticity field and interface dynamics, as the non-linear term tends to 
modify the vorticity field differently upon the sign of $B$. Thus, while for $B < 0$ non-linear terms tend to
increase the intensity
of vorticity production in the region $|x| < 1$ and decreasing it in the outer region $|x| > 1$, the opposite occurs for
$B > 0$. The direct consequence is that the two vorticity layers of opposite sign created at both sides of
the x-axis 
preferentially roll-up near the axis 
creating a jet for $B <0$ 
while for $B > 0$ the roll-up of the structure is induced in the outer region.

\section{Conclusion.}
\label{conclusion2D}

 \vspace{0.2cm}

 \noindent In this work, we studied the  production of vorticity at  an interface separating two immiscible incompressible fluids for a  two-dimensional flow.  We proposed a new decomposition of the vorticity flux which makes explicit its dependence on parameters such as surface tension $\sigma$,  viscosity $\mu$ and gravity $g$, the various factors are obtained by solving Laplace equations. In some   cases, in particular  the case $\rho^{(2)}/\rho^{(1)} <<1$ and $Re \gg 1$, it is possible to  solve analytically most of these Laplace equations and to reduce  the complexity of the procedure.  This approach can {\it a priori} be  extended to three-dimensional flows (we are currently working on this topic).

 \vspace{0.2cm}

 \noindent The case of   gravito-capillary wave  has been also  discussed based on this procedure. Analytical as well as numerical examples have been  presented.  From the  analytical side,  it leads to results already known but from  a new perspective. From the numerical perspective, it provides some quantitative predictions at short time that can be a good test for numerical codes or  enables a  qualitative understanding  of numerical results.

 \vspace{0.2cm}



 \vspace{0.2cm}

 \section*{Acknowledgements}    
 The authors would like to thank  L. Duchemin,  J.Magnaudet, S.Popinet and  S.Zaleski  for fruitful discussions.
\bibliographystyle{jfm}
  \bibliography{vorticity}


 \appendix

 \vspace{0.2cm}

\section{Numerical implementation for sources computations. }
\label{AppendixNumericalimplementation}

\noindent  To evaluate   the vorticity flux $\Sigma$, one should  compute  one  Poisson equation and  ten Laplace equations. The Poisson solver is easy to implement.  For  the five  discontinuous  fields    $ \Psi_{d \sigma}$,  $ \Psi_{d g} $,  $\Psi_{d \rho}$,   $\Psi_{d \mu}$, the boundary condition of    Dirichlet  type is imposed  on a boundary with a known geometry.  Practically, there exists  immersed boundary methods that solve numerically such a problem \citep{popinet2003gerris,selccuk2020fictitious}.\\

 \vspace{0.5cm}

\noindent  The remaining  five    fields   $  \Psi_{[[\rho]]}$,  $   \Psi_{\sigma}$,  $  \Psi_{g} $, $    \Psi_{\mu_m}$, $ \Psi_{ [[\mu]]}$   are continuous  across interface $(I)$ but the  normal derivative of which is discontinuous across $(I)$   of the form  $[[ \upsilon n_i\partial_i   \Phi ]]= G$ (see conditions \eqref{cond1}--\eqref{cond3}). These  five    fields  $\Psi_{\sigma}$,  $\Psi_{g} $,  $\Psi_{[[\rho]]}$  , $\Psi_{\mu_m}$, $\Psi_{ [[\mu]]}$   can  be obtained  by numerical methods. In a  volume of fluids approach,   one solves the Laplace  equation 
\begin{equation}
\Delta  \Psi = \nabla \cdot ( \nabla \Psi) = 0
\label{defrhomp1p2PmLAPLACEequOnLine}
 \end{equation}
 in each domain away from the cell crossed by the interface boundary.  For such cells   containing an interface,     special care is required so that  the derivative discontinuity is used in integrating this equation. Each such cell is subdivided into two sub-cells $\Omega_1$ and $\Omega_2$ occupied for   phase  $1$ and $2$ respectively. Since in each subcell, Laplace equation  is satisfied
\begin{equation}
 \upsilon^{(1)} \int_{\Omega_1} \nabla \cdot ( \nabla \Psi^{(1)}) d s_1=0,~~~~~ \upsilon^{(2)}  \int_{\Omega_2} \nabla \cdot ( \nabla \Psi^{(2)}) d s_2 =0
 \end{equation}
Now applying the divergence theorem in each phase and summing both expressions,  yield
\begin{equation}
  \upsilon^{(1)} \int_{S_1}  \vec{n}^{(out)} \cdot \nabla \Psi^{(1)}  dS_1+ \upsilon^{(2)} \int_{S_2} \vec{n}^{(out)}  \cdot   \nabla \Psi^{(2)} dS_2
  +\int_{(I)} [[\vec{n}^{(1\to 2)}    \cdot (\upsilon \nabla  \Psi) ]]  dS =0.
 \end{equation}
where the  closed surface of the   cell is  divided in  $S_1$  in phase 1 and  $S_2$  in phase 2 and $ n^{(out)} $ is the outward unit normal vector. In addition let us called  $(I)$ the portion of  interface cutting the cell.  For a quad/cube cell with face surface $\Delta S_f$ crossed by an interface of length $\Delta S_I$ we readily obtain
\begin{equation}
  \sum_f  \vec{n}^{(out)} \cdot  \bigg(\upsilon^{(1)}  \nabla \Psi^{(1)} \bigg)  c_f  +  
  \sum_f   \vec{n}^{(out)} \cdot \bigg(  \upsilon^{(2)}   \nabla \Psi^{(2)} \bigg) (1 - c_f)
  =- G \frac{\Delta S_I}{\Delta S_f}
 \end{equation}
 where $n^{1 \to 2}$ is the unit normal to the interface pointing   from fluid 1 to fluid 2,  $c_f$ is the face fraction of the  fluid 1  crossing a given face and 
 \begin{equation}
 G \equiv  [[ \upsilon \vec{n}^{(1\to 2)}\cdot \nabla \Psi]],
 \end{equation} 
 is given by   \eqref{cond1}--\eqref{cond2}--\eqref{cond3}. This numerical approach  is used in the computation of the different sources in section~\ref{bumpflowgrav} which assumes also continuity of function $\Psi$ crossing the interface. It is equivalent to
 solve the variable coefficient Poisson  equation 
\begin{equation}
 \nabla \cdot (\upsilon \nabla \Psi) = -G\delta({n}^{(1\to 2)})
 \label{defrhomp1p2PmPoissonequOnLine}
 \end{equation}

 
 \section{ Asymptotic case $A_{tw} \to 1$.}
 \label{vorticityproductionextremecaserho1rho2}

 \noindent  In this appendix,  we work in dimensionless variables and  study  functions $\Psi_{x}$ ($x=   \sigma, g, \rho, \rho_m,  \mu_m, [[\mu]]$) when fluid 1 is much lighter than fluid 2 i.e. $ {\rho^{(2)} }>> \rho^{(1)}$
 and  $\nu^{(2)} /\nu^{(1)} $  is of order one. A  small parameter  $ \epsilon \equiv \frac{ \upsilon^{(2)}}{ \upsilon^{(1)}}=\frac{ \rho^{(1)}}{ \rho^{(2)}} <<1$ exists.   Note  the relations $\epsilon=\frac{1-A_{tw}}{1+A_{tw}}  $   and
  \begin{equation}
 \frac{[[\upsilon]] \mu_m}{\nu_m}=-\epsilon \frac{\nu^{(1)}}{\nu^{(1)}+\nu^{(2)}} +
 \frac{\nu^{(1)}-\nu^{(2)}}{\nu^{(1)}+\nu^{(2)}} +\frac{1}{\epsilon}\frac{\nu^{(2)} }{\nu^{(1)}+\nu^{(2)}}
 \label{cond1expanbis1}
 \end{equation}
 \begin{equation}
 \frac{[[\upsilon]]~[[\mu]]}{\nu_m} =- \epsilon \frac{2\nu^{(1)}}{\nu^{(1)}+\nu^{(2)}} + 2 -\frac{1}{\epsilon}\frac{2 \nu^{(2)} }{\nu^{(1)}+\nu^{(2)}}
\label{cond1expanbis2}
 \end{equation}
 \begin{equation}
  \frac{\upsilon_m [[\mu]]}{\nu_m} =\epsilon \frac{\nu^{(1)}}{\nu^{(1)}+\nu^{(2)}}+\frac{\nu^{(1)}-\nu^{(2)}}{\nu^{(1)}+\nu^{(2)}}  -\frac{1}{\epsilon}\frac{\nu^{(2)} }{\nu^{(1)}+\nu^{(2)}}.
\label{cond1expanbis3}
 \end{equation}
When $A_{tw} \to 1$, $\epsilon \approx \frac{1-A_{tw}}{2} $

 \vspace{0.5cm} 
  
\noindent    First functions $\Psi_{d \sigma}$, $\Psi_{d g}$, $\Psi_{d \rho} $ do not depend on  the density ratio $\epsilon$ and all vary over  a characteristic   length scale  $L_0$.  Second the boundary condition   for $\Psi_{x}$  with $x=   \sigma, g, \rho$ reads 
 \begin{equation}
   n_i\partial_i    (\Psi^{(1)}_{ x}+ \Psi^{(1)}_{d x})    = \epsilon  n_i\partial_i  (\Psi^{(2)}_{x}+ \Psi^{(2)}_{d x})
 \end{equation}
Since  $\Psi^{(1)}_{d x}$ and $\Psi^{(2)}_{dx}$ are both harmonic  and $\Psi^{(1)}_{d x}=-\Psi^{(2)}_{dx}$   along the interface, these functions  must  vary   with the same characteristic length    $L_0$. This  leads to the simplification 
   \begin{equation}
   n_i\partial_i  (\Psi^{(1)}_{x}+ \Psi^{(1)}_{d x})   = 
 \epsilon n_i\partial_i  (\Psi^{(2)}_{x})
 \label{cond1bis1}
 \end{equation}
Similarly $\Psi^{(1)}_{x}$ and $\Psi^{(2)}_{x}$ are   harmonic  and  $\Psi^{(1)}_{x}=\Psi^{(2)}_{x}$  on the interface:  these functions    vary   with the same characteristic length hence a further simplification
 \begin{equation}
   n_i\partial_i  (\Psi^{(1)}_{x}+ \Psi^{(1)}_{d x})   = 0,~~~\hbox{for }~~~x=   \sigma, g, \rho
 \label{cond1bis2}
 \end{equation}
Fields $ \Psi_{\sigma}$, $ \Psi_{g}$, $\Psi_{ [[\rho]]}$ hence verify continuity   across the interface and 
the above simplified conditions. It is easily seen that  this leads to 
\begin{equation}
\Psi_{x}^{(1)}  = -    \Psi^{(1)}_{d x},~~~~\Psi_{x}^{(2)}  = \Psi^{(2)}_{d x} ~~~\hbox{for }~~~x=   \sigma, g, \rho
 \end{equation}
Because of conditions  \eqref{LaplaceBCdiscontnew}, this imposes at the interface
 \begin{equation}
 \Psi_{\sigma} = \frac{ \kappa}{2},~~~ \Psi_{g} = -\frac{\eta}{2},~~~~ \Psi_{ [[\rho]]} =\frac{1}{2}{\Psi_{\rho_m}},
  \label{LaplaceBCdiscontnewAA}
\end{equation}
 It is thus not necessary  to solve Laplace equations for $\Psi_{\sigma}$, $\Psi_{g} $, or  $\Psi_{ [[\rho]]} $ in such an approximation. On the interface, this yields  
\begin{equation}
  \hat \Psi_{\Sigma} = - \frac{1}{2} \frac{ \kappa}{We}  -   Fr~\hat \eta + \hat \Psi_{\rho_m} + \Psi_{\rm{visc}}
   \label{Sigmasource1nodim}
\end{equation}
where the term $\Psi_{\rm{visc}}$ is related to viscous effects discussed below.

 \vspace{0.2cm}

\noindent  In addition,   $\Psi_{ \mu_m}$  satisfies Laplace equation and 
  \begin{equation}
   n_i\partial_i    (\Psi^{(1)}_{\mu_m})- \epsilon n_i\partial_i    (\Psi^{(2)}_{\mu_m})=
    n_i\partial_j \partial_j  u^{(1)}_i -  \epsilon n_i\partial_j \partial_j  u^{(2)}_i
\label{cond2bis12}
 \end{equation} 
Since $ \Psi_{\mu_m}$ is a harmonic function and continuous across the surface then one may neglect the second l.h.s. term
 \begin{equation}
   n_i\partial_i    (\Psi^{(1)}_{\mu_m}) =
    n_i\partial_j \partial_j  u^{(1)}_i -  \epsilon   n_i\partial_j \partial_j  u^{(2)}_i
\label{cond2bis13}
 \end{equation}

\noindent { \bf Condition  $Re>>1$}

 \vspace{0.5cm} 

\noindent  When $Re>>1$  and $\nu^{(2)}/\nu^{(1)} =O(1)$,  a boundary layer is  present of dimensionless size $\delta^{(r)} =  \frac{\nu^{(r)}}{\nu_m} \sqrt{Re}$ in each phase and  of comparable width. It is however a weak boundary layer since vorticity is of order one contrary to the boundary layer on a solid. Hence quantity $n_i\partial_j \partial_j  u^{(r)}_i $ could be of order $O(\sqrt{Re})$ or less for both phases and the second term
 of the r.h.s. of equation \eqref{cond2bis13} may be again neglected compared to the first of the r.h.s. 
 \begin{equation}
   n_i\partial_i    (\Psi^{(1)}_{\mu_m}) =
    n_i\partial_j \partial_j  u^{(1)}_i 
\label{cond2bis14}
 \end{equation} 
In that approximation, one only solves  the  Laplace equation   in the phase 1 with the above Neumann condition. Furthermore   we   need $\Psi^{(2)}_{ \mu_m}$  on the interface. This value is given by    the Dirichlet condition $\Psi^{(2)}_{ \mu_m}     =   \Psi^{(1)}_{ \mu_m}$.  The field  $\Psi_{ d \mu}$  satisfies Laplace equation and 
  \begin{equation}
  \Psi^{(1)}_{ d \mu}     =  -\Psi^{(2)}_{ d \mu}= \bigg( \kappa \vec{u} \cdot \vec{n}
  - \, \vec{t} \cdot \vec{\nabla} (\vec{u} \cdot \vec{t})\bigg) 
\label{cond2bis11}
 \end{equation}
 It is of order $O(1)$.  Finally function $\Psi_{ [[\mu]]}+ \Psi_{d \mu}$ is  harmonic, and satisfies along the interface
\begin{equation}
  n_i\partial_i   ( \Psi^{(1)}_{ [[\mu]]}+ \Psi^{(1)}_{d \mu})  - \epsilon n_i\partial_i   ( \Psi^{(2)}_{ [[\mu]]}+ \Psi^{(2)}_{d \mu})    = \frac{1}{2} \bigg( n_i \partial_j \partial_j  u^{(1)}_{i}  +  \epsilon  n_i\partial_j \partial_j    u^{(2)}_{i}  \bigg) 
\label{cond31}
 \end{equation}
 One neglects the second r.h.s term as above  if $\nu^{(2)}/\nu^{(1)} =O(1)$. In addition, because of its continuity, $\Psi_{ [[\mu]]}$ varies along the interface in a similar manner in both domain  so that   one may neglect the   second l.h.s. term
\begin{equation}
n_i\partial_i    \Psi^{(1)}_{ [[\mu]]}  =-  n_i\partial_i  \big(\Psi^{(1)}_{d \mu} - \frac{1}{2}\Psi^{(1)}_{ \mu_m}]\big)
\label{cond3bisbis}
 \end{equation}
This is solved in domain fluid 1 
 \begin{equation}
 \Psi^{(1)}_{ [[\mu]]}  = -\Psi^{(1)}_{d \mu} + \frac{1}{2}\Psi^{(1)}_{ \mu_m}.
 \label{phimudiffrel1A}
  \end{equation} 
The   continuity  of  $\Psi_{ [[\mu]]}$ across the interface then implies
 \begin{equation}
 \Psi^{(2)}_{ [[\mu]]}  =  \Psi^{(2)}_{d \mu} + \frac{1}{2}\Psi^{(2)}_{ \mu_m}.
 \label{phimudiffrel2A}
  \end{equation} 
\noindent  For   $Re >>1$,  $ \hat  \Psi_{\mu_m} $,  $ \hat   \Psi_{[[\mu]]} $ are of order $O(\sqrt{Re})$.
Using expansions \eqref{cond1expanbis1}, \eqref{cond1expanbis2}, \eqref{cond1expanbis3} and  relation \eqref{phimudiffrel1A}, the  term $\Psi_{\rm{visc}} $ in  equation \eqref {Sigmasource2}  is of order $O(1/\sqrt{Re}$)  and 
 \begin{equation}
\Psi_{\Sigma} \approx   - \frac{1}{2} \frac{ \kappa}{We}  -   Fr~\hat \eta +    \hat \Psi_{\rho_m}~~~\hbox{for}~~~Re>>1
\end{equation}
 
 \vspace{0.5cm}

\noindent { \bf Stokes  condition   $Re<<1$}

 \vspace{0.5cm} 
 
When $Re<<1$  and $\nu^{(2)}/\nu^{(1)} =O(1)$,  the quantity $n_i\partial_j \partial_j  u^{(r)}_i $ is of the same order   for both phases  and the second term
 in the r.h.s. may be hence  again neglected.
 \begin{equation}
   n_i\partial_i    (\Psi^{(1)}_{\mu_m}) =
    n_i\partial_j \partial_j  u^{(1)}_i 
\label{cond2bis14bis}
 \end{equation} 
In that approximation, one only solves  the  Laplace equation   in the phase 1,  with the Neumann condition
\begin{equation}
 n_i\partial_i  \Psi^{(1)}_{ \mu_m}     =  n_i\partial_j \partial_j   u^{(1)}_i
\label{cond3bis}
 \end{equation}
 which depends on phase 1 only. Furthermore   we   need $\Psi^{(2)}_{ \mu_m}$  on the interface. This value is given by    the Dirichlet condition $\Psi^{(2)}_{ \mu_m}     =   \Psi^{(1)}_{ \mu_m}$.  The field $\Psi_{ d \mu}$  satisfies Laplace equation and 
  \begin{equation}
  \Psi^{(1)}_{ d \mu}     =  -\Psi^{(2)}_{ d \mu}= \bigg( \kappa \vec{u} \cdot \vec{n}
  - \, \vec{t} \cdot \vec{\nabla} (\vec{u} \cdot \vec{t})\bigg) 
\label{cond2bis11}
 \end{equation}
 Finally function $\Psi_{ [[\mu]]}+ \Psi_{d \mu}$ is  harmonic, and satisfies along the interface
\begin{equation}
  n_i\partial_i   ( \Psi^{(1)}_{ [[\mu]]}+ \Psi^{(1)}_{d \mu})  - \epsilon n_i\partial_i   ( \Psi^{(2)}_{ [[\mu]]}+ \Psi^{(2)}_{d \mu})    = \frac{1}{2} \bigg( n_i \partial_j \partial_j  u^{(1)}_{i}  +  \epsilon  n_i\partial_j \partial_j    u^{(2)}_{i}  \bigg) 
\label{cond31}
 \end{equation}
 One neglects the second r.h.s term as above  if $\nu^{(2)}/\nu^{(1)} =O(1)$. In addition, because of its continuity, $\Psi_{ [[\mu]]}$ varies along the interface in a similar manner in both domain  so that   one may neglect the   second l.h.s. term
\begin{equation}
n_i\partial_i    \Psi^{(1)}_{ [[\mu]]}  =-  n_i\partial_i  \big(\Psi^{(1)}_{d \mu} - \frac{1}{2}\Psi^{(1)}_{ \mu_m}]\big)
\label{cond3bisbis}
 \end{equation}
This is solved in domain fluid 1 
 \begin{equation}
 \Psi^{(1)}_{ [[\mu]]}  = -\Psi^{(1)}_{d \mu} + \frac{1}{2}\Psi^{(1)}_{ \mu_m}.
 \label{phimudiffrel1}
  \end{equation} 
The   continuity  of  $\Psi_{ [[\mu]]}$ across the interface then implies
 \begin{equation}
 \Psi^{(2)}_{ [[\mu]]}  =  \Psi^{(2)}_{d \mu} + \frac{1}{2}\Psi^{(2)}_{ \mu_m}.
 \label{phimudiffrel2}
  \end{equation} 
  As a consequence on the interface at order zero
  \begin{equation}
 \Psi_{ [[\mu]]}    = -\bigg( \kappa \vec{u} \cdot \vec{n}  - \, \vec{t} \cdot \vec{\nabla} (\vec{u} \cdot \vec{t})\bigg)   + \frac{1}{2}\Psi_{ \mu_m}.
 \label{phimudiffrelinter}
  \end{equation}  
Replacing these expressions into the full vorticity source we readily find that the source is zero at leading order
and therefore it is required to obtain the functions at the next order, requiring
to evaluate them numerically in a general case.
As a conclusion, although for $Re<< 1$ 
the viscous terms are always preponderant and control the vorticity production
$$\hat \Psi_{\Sigma} \approx \Psi_2.$$
Note that there is no obvious advantage
between computing 
the first order approximation and the full expression for Eq. \ref{Sigmasource2}.\\


\section{Source field   for viscous capillary gravity waves.}
\label{Viscoussourcefieldgravitywave}

\noindent  We start by expressing   coefficients $A^{(1)}$, $ B^{(1)}$,  $A^{(2)}$, $B^{(2)}$ as a function of $\eta_0$. 
Note that  the tangential velocity field is continuous across the interface. After linearization this implies
\begin{equation}
ik A^{(1)} +\kappa^{(1)} B^{(1)}  = ikA^{(2)} - \kappa^{(2)} B^{(2)} 
\label{eq:Vinterface2BIS}
\end{equation}
The linearized kinematic condition at the interface 
\begin{equation}
\partial_t \eta = \partial_{y}\phi^{(r)}(x,y=0,t) + \partial_{x}\psi^{(r)}(x,y=0,t),~~~~r=1,2;
\label{eq:Vinterface}
\end{equation}
yields  two supplementary relations 
 \begin{equation}
 B^{(1)}   =      - \frac{\varpi}{k} \eta_0 - i \frac{|k|}{k} A^{(1)};~~~~~
 B^{(2)}   =      - \frac{\varpi}{k} \eta_0 + i \frac{|k|}{k} A^{(2)} 
 \label{eq:Vinterface1new}
 \end{equation}
Finally the jump on vorticity in its linearized form 
\begin{equation}  
[[ \mu {\omega}_z]]=-[[2\mu]] \frac{\partial u_y}{\partial  x}=- i k [[2\mu]]    u_y 
\label{MeqnNormalParallelvorticitebislinea}
\end{equation}
yields  the fourth equation 
\begin{equation}  
 \mu^{(1)}  \big(2ik|k|  A^{(1)} +[k^2 +(\kappa^{(1)})^2]B^{(1)}\big)= \mu^{(2)}  \big( - 2ik|k|  A^{(2)} +[k^2 +(\kappa^{(2)})^2 ]B^{(2)}\big)
\label{MeqnNormalParallelvorticitebislinea1}
\end{equation}
The solution of the system 
yields
\begin{equation}  
 |k| A^{(1)}=   i \varpi \eta_0  - 2k \frac{i \varpi  \rho^{(2)}  + k[[\mu]] b^{(2)} }{ \rho^{(2)}b^{(1)} +\rho^{(1)} b^{(2)}}\eta_0,
 \label{MeqnNormalParallelvorticitebislinea2}
\end{equation}
\begin{equation}  
 |k| A^{(2)}=  -i\varpi \eta_0+ 2k  \frac{i \varpi \rho^{(1)} - k[[\mu]] b^{(1)} }{ \rho^{(2)}b^{(1)} +\rho^{(1)} b^{(2)}}\eta_0,
\label{MeqnNormalParallelvorticitebislinea2BIS}
\end{equation}

\noindent   Let us now compute $\Sigma$  using the source terms  in  \eqref{MeqnSOmegaInterfaceLINEARVISCOUS} i.e.  $ \Psi_{\sigma}$, $\Psi_{g}$, $\Psi_{\mu_m}$, $\Psi_{[[\mu]]}$ and ${u}_x$. It is easy to understand that  a field $\Psi_{q} $ where $q$ is  selected among one of sources ${d \sigma}$, ${d g}$, ${d \mu}$, ${\sigma}$, ${g}$,  ${\mu_m}$ or ${[[\mu]]}$, satisfies a Laplace equation within the two fluid phases. As the consequence, this imposes 
\begin{equation}
  \Psi_{q} =
    \begin{cases}
     A^{(1)}_{q} \exp { \bigg(i (kx-\varpi(k) t)   -  \mid k \mid y \bigg)} & \text{if  $0 < y$}\\
      A^{(2)}_q  \exp { \bigg(i (kx-\varpi(k) t)   +  \mid k \mid y \bigg)}  & \text{if $y\le 0$}
    \end{cases}   
    \label{equationPSIQ}    
\end{equation}
Replacing these values in equation  \eqref{MeqnSOmegaInterfaceLINEARVISCOUS}   yields       
$\Sigma = ik  \Psi_{\Sigma}  \exp { i (kx-\varpi(k) t) }$  with
\begin{equation}
 {\Psi_{\Sigma}}_0 = \bigg( -\sigma \upsilon_m k^2  \eta_0
 +   [[\upsilon]]\big( \sigma  A_{\sigma}   
+ g [[\rho]]    A_{g}      
 -     \rho_m g \eta_0 \big) \bigg)
  +   [[\upsilon]]  \mu_m    A_{\mu_m}
 +    [[\mu]] \bigg( - \,2  \upsilon_m\,  ik  u_x
+ [[\upsilon]]  A_{[[\mu]]}     
  \bigg)
 \label{MeqnAppendixSOmegaInterfaceVISCOUS}
\end{equation} 
For infinitesimal amplitudes,   the  boundary conditions  for the various  $\Psi_q$  fields should be linearized   at   $y=0$.  By matching these conditions,  the constants of   fields  $\Psi_{d g}$, $\Psi_{g}$, $\Psi_{d \sigma}$ and $\Psi_{\sigma}$  in the Laplace  equations \eqref{equationPSIQ}  are  found.  
  
 \vspace{0.5cm}

 \noindent The  boundary conditions for   fields  $\Psi_{d g}$  and $\Psi_{d \sigma}$   linearized  at   $y=0$ read  
 \begin{equation}
\Psi^{(1)}_{d g}(y=0,t)=-\Psi^{(2)}_{d g}(y=0,t)=\frac{\eta_0}{2},~~~~\Psi^{(1)}_{d \sigma}(y=0,t)=-\Psi^{(2)}_{d \sigma}(y=0,t)=-k^2 \frac{\eta_0}{2}.
\label{LaplacePeBCLinear}
\end{equation}
This implies  that $\Psi_{d \sigma} =  - k^2 \Psi_{d g}$.  The Laplace equation    with the above conditions leads   to
\begin{equation}
  A^{(1)}_{d g} = \frac{\eta_0}{2},~~~~~~A^{(2)}_{d g}= - \frac{\eta_0}{2}      
\end{equation}
Thereafter one introduces these expressions in the linearized boundary conditions of 
$\Psi_{g}$    yielding
\begin{equation}
  A_{g} = -A_{tw} \frac{\eta_0}{2} ,~~~~\Psi_{\sigma} =- k^2 \Psi_{g}.
\end{equation}
   \noindent  The source can be thus simplified
\begin{equation}
 {\Psi_{\Sigma}}_0 = -2  \frac{\varpi^2_{inv}(k)}{|k|}   \eta_0  +   [[\upsilon]]  \mu_m    A_{\mu_m}
 +    [[\mu]] \bigg(-2  \upsilon_m\,  ik  u_x+ [[\upsilon]]  A_{[[\mu]]}       \bigg)
 \label{MeqnAppendixSOmegaInterfaceVISCOUSsimplifiedappend000}
\end{equation}  
 
   \vspace{0.5cm}

 \noindent The    field   $\Psi_{\mu_m}$ satisfies the linearized version of  continuity and  condition \eqref{cond2} 
\begin{equation}
  \Psi^{(1)}_{\mu_m}(y=0) = \Psi^{(2)}_{\mu_m}(y=0) ~~~\hbox{and}~~~[[\upsilon \frac{\partial}{\partial y}  \Psi_{ \mu_m}]]     =  [[  \partial_j \partial_j  ( \upsilon u_y)]] 
  \label{LaplaceBCdiscontnewViscousmum}
\end{equation}
The two  conditions reads 
 \begin{equation}
A_{\mu_m}=\frac{ \varpi^2}{2 |k|\upsilon_m }   \bigg( \frac{1 }{ \mu^{(1)} } - \frac{1 }{ \mu^{(2)} }\bigg)\eta_0   +  i  \frac{\varpi }{2\upsilon_m }  \bigg( \frac{1 }{ \mu^{(1)} } A^{(1)}+  \frac{1 }{ \mu^{(2)} }A^{(2)}  
  \bigg)
  \label{LaplaceBCdiscontnewViscous1mumbis}
\end{equation}

  \vspace{0.5cm}

 \noindent The    field      $\Psi_{d \mu}$ satisfies  the Dirichlet condition \eqref{LaplaceBCdiscontnew} which once linearized, imposes
\begin{equation}
 \Psi^{(1)}_{d \mu}(y=0) = -\Psi^{(2)}_{d \mu}(y=0)=  - \, ik  {u}_x(y=0)
  \label{LaplaceBCdiscontnewViscous1}
\end{equation} 
 \noindent  The continuity \eqref{eq:Vinterface2BIS}  of velocity component $u_x$  across the interface   imposes
\begin{equation}
 A^{(1)}_{d \mu} =(k^2  -  \kappa^{(1)}|k|)A^{(1)}+ i \kappa^{(1)} \varpi \eta_0 
 \end{equation} 
\begin{equation}
A^{(2)}_{d \mu}  =-(k^2- \kappa^{(2)}   |k|) A^{(2)}  +i  \kappa^{(2)} \varpi  \eta_0
\end{equation}

 \vspace{0.5cm}

 \noindent The    field     $\Psi_{[[\mu]]}$   satisfies    continuity and   condition   \eqref{cond3}   linearized  across the interface.
The first condition leads to  $\Psi^{(1)} (y=0)=\Psi^{(2)}(y=0)$  that is
$A^{(1)}_{[[\mu]]}   =A^{(2)}_{[[\mu]]}$. The second condition at $y=0$  yields 
\begin{equation}
 [[\upsilon \frac{\partial}{\partial y}  \Psi_{[[\mu]]}]]  = - [[ \upsilon \frac{\partial}{\partial y}  \Psi_{d \mu}]] 
+\frac{1}{2} \bigg(   \partial_j \partial_j  ( \upsilon^{(1)}  u^{(1)}_{y}) +    \partial_j \partial_j  (\upsilon^{(2)}  u^{(2)}_{y})  \bigg) 
\label{cond3linCrochetmu}
 \end{equation}
which can be rewritten as 
 \begin{equation}
 A_{[[\mu]]}     = -A_{tw}     A^{(1)}_{d \mu}  +A'_{[[\mu]]},
 \label{cond3linCrochetmuBIS}
 \end{equation}
  \begin{equation}
 A'_{[[\mu]]}     =  
 \frac{1}{2|k|} \bigg( \frac{1 }{ \mu^{(1)} }  + \frac{1 }{ \mu^{(2)} }  \bigg)   \frac{\varpi^2 }{2\upsilon_m }\eta_0  
 + i  \frac{\varpi }{4\upsilon_m }   \bigg(\frac{1 }{ \mu^{(1)} } A^{(1)}  - \frac{1 }{ \mu^{(2)} }   A^{(2)}  \bigg). 
\label{cond3linCrochetmuBIS}
 \end{equation}
 
 \vspace{0.5cm} 
 
 By summing these various sources, the total source becomes after some algebraic manipulations 
  \begin{equation}
 \Psi_{\Sigma}=   \alpha_0 \eta_0  +  \alpha_1 A^{(1)}  + \alpha_2 A^{(2)}
  \label{MeqnAppendixSOmegaInterfaceVISCOUSsimplifiedappend001}
\end{equation}   
  \begin{equation}
 \alpha_0 = -2  \frac{\varpi^2_{inv}}{|k|}   
 +\frac{[[\mu]]}{\rho_m}  i (\kappa^{(1)}-\kappa^{(2)}) \varpi        
 \label{MeqnAppendixSOmegaInterfaceVISCOUSsimplifiedappend0}
\end{equation}
 \begin{equation}
 \alpha_1 = i A_{tw} \varpi   +\frac{ [[\mu]]}{\rho_m} (k^2  -  \kappa^{(1)}|k|)      
 \label{MeqnAppendixSOmegaInterfaceVISCOUSsimplifiedappend1}
\end{equation}  
 \begin{equation}
  \alpha_2 =   i A_{tw} \varpi    +\frac{[[\mu]]}{\rho_m}   (k^2  -  \kappa^{(2)}|k|)       
 \label{MeqnAppendixSOmegaInterfaceVISCOUSsimplifiedappend2}
\end{equation}  

where the values of coefficients $A^{(1)}$ and $A^{(2)}$ are given in Eqs. \ref{MeqnNormalParallelvorticitebislinea2}-\ref{MeqnNormalParallelvorticitebislinea2BIS}.

\section{Computations near time $t=0$ for viscous gravity waves.}
\label{RTComput}

 \vspace{0.2cm}
     
 \noindent Here we work in dimensionless units. It is recalled that the curvilinear variable $s$ increases along  $\vec{t}^{1\to 2}$ and the orthogonal variable $s_{\perp}$ increases along $\vec{n}^{1\to 2}$ in the  Frenet-Serret frame. Let us  evaluate the circulation per unit length produced during the time period near time $t=0$ in each monophasic domain 
 \begin{equation}
\gamma^{(1)} =\int^{0}_{-\infty}\omega^{(1)}\, \text{d}s_{\perp}~~~~~
\gamma^{(2)} =\int^{\infty}_{0}\omega^{(2)}\, \text{d}s_{\perp}
\label{circulationmonophasicbisappend}
\end{equation}
In that period,  the fluid is almost at rest and  vorticity is zero initially. As a consequence equations \eqref{MeqnIntegomegaJijV1}  read
 \begin{equation}
\frac{\partial}{\partial t}\left(\int_{A^{(r)}}\omega\,\text{d}x  \text{d}y\right)
  =    -\int_{(C_r) } J^{(r)}_{j}\,n_{j}\,\text{d}s_c + \int_{(I) }  \Sigma^{(r)}\,\text{d}s,~~~r=1,2
  \label{MeqnIntegomegaJijV1linearappend}
\end{equation}
where  the loop is the union of   $(I)$ a stretch   $ds$  along the interface and   $(C_r) $  made of two lines along the $s_{\perp} $-axis in fluid $r$  closing at infinity.

\subsection{Computations near time $t=0$ discarding diffusion along the interface.}
 
 The r.h.s term becomes non zero and    provides in each phase
 \begin{equation}
\frac{\partial \gamma^{(1)} }{\partial t}     =     \frac{1+r_{\rho}}{2} \frac{1}{Re}  \frac{\partial^2 \gamma^{(1)} }{\partial s^2}   +      {\Sigma}^{(1)}(s,0),     
   \label{MeqnIntegomegaJijVtotbisLinearizedBISfiniteampliappend1}
 \end{equation}
 \begin{equation}
\frac{\partial \gamma^{(2)} }{\partial t}     =     \frac{1+r_{\rho}}{2r_{\rho}} \frac{1}{Re}  \frac{\partial^2 \gamma^{(2)} }{\partial s^2}   +      {\Sigma}^{(2)}(s,0).     
   \label{MeqnIntegomegaJijVtotbisLinearizedBISfiniteampliappend2}
 \end{equation}
 When discarding diffusion along the interface, one obtains  near time $t=0$
  \begin{equation}
  \gamma^{(r)}(s,t)    =    {\Sigma}^{(r)}(s,0)   ~t   ,~~~~r=1,2
   \label{MeqnIntegomegaJijVtotbisLinearizedBISfiniteampli}
 \end{equation}
This implies that the circulation in phase $r$ in the half plane $x \ge 0$ evolves according to 
\begin{equation}
  \Gamma^{(r)}_{x \ge 0}(t)    =    \bigg( \int_{(I),~x \ge 0}   {\Sigma}^{(r)}(s,0)\, ds \bigg)~t,~~~~r=1,2
   \label{MeqnIntegomegaJijVtotbisLinearizedBISfiniteampli}
 \end{equation}

 \vspace{0.2cm}

 \noindent To go a step further, we may    evaluate the vorticity produced   during the first instants. Since the velocity field is almost zero, equation \eqref{MeqVorticity} implies that vorticity obeys a pure diffusion equation in the normal direction to the interface at any point of the interface  with a Neumann  boundary condition at the interface which is nothing but equation~\eqref{MeqnIntegomegaJijVsigma}~: for $s_{\perp}\le 0$
 \begin{equation}
  \partial_t \omega^{(1)}(s,t)    =  \frac{1+r_{\rho}}{2} \frac{1}{Re} \partial^2_{s_{\perp}} \omega^{(1)},~~~~ \frac{1+r_{\rho}}{2} \frac{1}{Re} \partial_{s_{\perp}} \omega= \Sigma^{(1)}_A(s,t=0) 
   \label{MeqnIntegomegadifusionphase1}
 \end{equation}
 for $s_{\perp}\ge 0$
  \begin{equation}
  \partial_t \omega^{(2)}(s,t)    =  \frac{1+r_{\rho}}{2r_{\rho}} \frac{1}{Re} \partial^2_{s_{\perp}} \omega^{(2)},~~~~~~ -\frac{1+r_{\rho}}{2r_{\rho}} \frac{1}{Re} \partial_{s_{\perp}} \omega= \Sigma^{(2)}_A(s,t=0) 
   \label{MeqnIntegomegadifusionphase2}
 \end{equation}
The solution of these two  equations are  known to be 
 \begin{equation}
 \omega(s,s_{\perp},t) =   
    \begin{cases}
     \omega^{(1)}(s,s_{\perp},t)  =  - 4\frac{  \Sigma^{(1)}_A(s,t=0)}{\sqrt{2}\delta^{(1)}}  G (-\frac{s_{\perp}}{\sqrt{2}\delta^{(1)}})~t~~& \text{if  $ s_{\perp} \le 0$}\\
       \omega^{(2)}(s,s_{\perp},t) =   -4 \frac{    \Sigma^{(2)}_A(s,t=0) }{\sqrt{2}\delta^{(2)}}  G (\frac{s_{\perp}}{\sqrt{2}\delta^{(2)}})~t~~ & \text{if $s_{\perp} \ge 0  $}
    \end{cases}       
\label{MeqinitBumpinterface1append}
\end{equation}
with
$$
 G(x) \equiv  \int^{x}_{0}  \bigg[1- Erf(x) \bigg]d x'  - \frac{1}{\sqrt{\pi}} =  -\frac{1}{\sqrt{\pi}}   \exp (-x^2)    + x\bigg[1- Erf(x) \bigg]
$$
and  
\begin{equation}
 \delta^{(1)}=    \sqrt{1+r_{\rho}} \sqrt{\frac{t}{Re}},~~~~~
 \delta^{(2)}=      \sqrt{\frac{1+r_{\rho}}{ r_{\rho}} }\sqrt{\frac{t}{Re}}
\label{MeqinitBumpinterface2append}
\end{equation}
Since  there is no jump of vorticity at interface because $[[\mu]]=0$  and since 
$$
  \hbox{when}~~~~x ~\to~~ 0,~~~ G(x) ~\to~  -\frac{1}{\sqrt{\pi}},~~~~~~~~\int^{\infty}_{0} G (z) dz=  -\frac{1}{4}
$$
It is easily found that
$$
\Sigma^{(1)}(s,t=0) = \sqrt{ r_{\rho} } \Sigma^{(2)}(s,t=0) 
$$
Since   $\Sigma=\Sigma^{(1)}+\Sigma^{(2)} $
\begin{equation}
\Sigma^{(1)}(s,t=0)=\frac{\sqrt{r_{\rho}}}{1+\sqrt{r_{\rho}}}   \Sigma(s,t=0), ~~
\Sigma^{(2)}(s,t=0)=\frac{1}{1+\sqrt{r_{\rho}}}  \Sigma(s,t=0)
\label{initBumpinterface3append}
\end{equation}
Using these expressions, the  circulation in each phase for $x\ge 0$ evolves according to
\begin{equation}
\Gamma^{(1)}_{x \ge 0}=\sqrt{r_{\rho}} \Gamma^{(2)}_{x \ge 0},~~~~~\Gamma^{(2)}_{x \ge 0}=\frac{1}{(1+\sqrt{r_{\rho}})}  \bigg( \int_{(I),~x \ge 0}{\Sigma}(s,0)\, ds \bigg)~t, 
   \label{MeqnIntegomegaJijVtotbisLinearizedBISfiniteampliBISappend}
 \end{equation}
and  enstrophy $E$ in the whole domain  according to 
\begin{equation}
E = \int \int \omega^2 dx dy=\int \int \omega^2 ds ds_{\perp} =I(r_{\rho}) \bigg( \int (  \Sigma)^2 ds\bigg) ~\sqrt{Re}~t^{3/2} 
\label{initBumpenstroappend}
\end{equation}
with
\begin{equation}
I(r_{\rho})=\frac{ 16}{\sqrt{2}}  \frac{\sqrt{2}-1}{3\sqrt{\pi}} \sqrt{r_{\rho} \over  1 + r_{\rho}} \frac{1}{(1 + \sqrt{r_{\rho}})}
\label{initBumpenstro2append}
\end{equation}
where one uses
 $$
 \int^{\infty}_{0} G^2 (z) dz=  \frac{\sqrt{2}-1}{3\sqrt{\pi}}.
$$

    \end{document}